\documentclass[aps,floatfix]{revtex4-2}
\usepackage{epsfig}
\usepackage{color}
\usepackage{amssymb}
\usepackage{amsmath}
\usepackage[ngerman,english]{babel}
\usepackage{multirow}
\usepackage{graphicx}
\usepackage{subfigure}
\usepackage{ulem}

\newcommand{\be}{\begin{equation}}
\newcommand{\ee}{\end{equation}}
\newcommand{\bea}{\begin{eqnarray}}
\newcommand{\eea}{\end{eqnarray}}

\newcommand{\GR}[1]{\textcolor{red}{#1}}
\newcommand{\HR}[1]{\textcolor{blue}{#1}}

\begin{document}

\title{The virial expansion of the Hydrogen equation of state in comparison to PIMC simulations: the quasiparticle concept, IPD, and  ionization degree }
\author{Gerd R\"opke$^{1}$}
\author{Chengliang Lin$^{2}$}
\author{ W. Ebeling$^{3}$}
\author{H. Reinholz$^{1}$}
\affiliation{
$^1$Institute of Physics, University of Rostock, Rostock, Germany\\
$^2$ National Key Laboratory of Computational Physics, Institute of Applied Physics and Computational Mathematics, Beijing, People’s Republic of China\\
$^3$Institute of Physics, Humboldt University Berlin, Germany}

\email{gerd.roepke@uni-rostock.de}

\date{\today}
\begin{abstract}
The properties of plasmas in the low-density limit are described by virial expansions. Analytical expressions are known for the lowest virial coefficients from Green's function approaches.
Recently, accurate path-integral Monte Carlo (PIMC) simulations were performed for the hydrogen plasma at low densities by Filinov and Bonitz [Phys. Rev. E {\bf 108}, 055212 (2023)], which made a comparison of the virial expansions and the derivation of interpolation formulas possible.
The exact expression for the second virial coefficient is used to test the accuracy of the PIMC simulations and the range of application of the virial expansions.
To describe plasmas in a wider range of density and temperature, the concept of quasiparticles is considered.
Medium modifications of free and bound states are obtained from the spectral function.
Mean-field effects are presented, such as exchange terms, Pauli blocking and screening. 
The density expansions of the quasiparticle shifts is considered.
The combination of PIMC simulations with benchmarks from exact virial expansion results allows us to obtain precise results for the EoS in the low-density range.
At low densities, the results are compared with the Saha equation to introduce the medium-dependent ionization potential.
The relation to the Beth-Uhlenbeck formula and concepts such as the Mott effect, ionization potential depression (IPD), and ionization degree are discussed.
The limits of current PIMC results for hydrogen plasmas are shown.
Further improvements of the PIMC simulations are required to compare with analytical benchmarks.
\end{abstract}

\maketitle

\section{Introduction}\label{sec0}

Hydrogen is the simplest and also the most distributed plasma. Therefore, accurate knowledge of the properties of hydrogen is a main request.
Thermodynamic, transport and optical properties have been investigated. The composition of a plasma, in particular the ionization degree, is of essential importance for properties such as the electrical conductivity and spectroscopic properties, for instance in astrophysics.

To describe the properties of hydrogen plasmas in a wide range of thermodynamic parameters, various theoretical approaches have been worked out, which treat the Coulomb interaction and the corresponding correlation effects. The review "Toward first principles-based simulations of dense hydrogen"  by Bonitz et al. \cite{Bonitz24} represents the current state of research in this field.  We will now give a short summary of those approaches which are relevant for our discussion of the equation of state (EoS).

Within the framework of quantum statistics, analytical approaches have been worked out using the method of thermodynamic Green's functions and Feynman diagrams \cite{KKER86}. 
Since these approaches are based on a perturbation expansion, exact results are obtained only in some limiting cases such as low densities, high temperatures and high densities. For a further extension of this approach's applicability,  the virial expansions of the equation of state (EoS) of hydrogen plasmas will be extensively analysed in the 
present work.

So-called {\it ab initio} calculations have been performed within the density-functional theory-molecular dynamics (DFT-MD) simulations, approximating the exchange-correlation effects of the electrons by an appropriate energy-density functional.
A current presentation of these approaches is found in the reviews \cite{Bonitz24,Mandy20}, where further references are given.
The DFT-MD simulations are very successful to describe the H plasma at high densities, in particular in the range where the electrons are degenerate so that the approximate treatment of electron-electron ($e-e$) interaction is less important. 
However, this approach fails to obtain the rigorous results in the low-density limit where $e-e$ correlations become relevant. 

A rigorous treatment of $e - e$ correlations is performed with  path-integral Monte-Carlo (PIMC) simulations,
but has shortcomings such as the small number of particles in the simulation box and the sign problem. 
Thanks to the increasing computer capabilities,
PIMC simulations can provide us with highly accurate values for the EoS.
For hydrogen, tables have been published by Filinov and Bonitz \cite{Filinov23} that give the pressure $p$ and the internal energy $E/N$ as function of temperature $T$ and electron density $n$ with relative statistical errors in the range of $10^{-4}$. 
The new simulations \cite{Filinov23} solved the finite size problem with high accuracy but suffer from the sign problem so that results are obtained only for particle densities $n$, for which the dimensionless parameter $r_s=\left[3/(4 \pi n a_{\rm B}^3)\right]^{1/3}$ is above the critical value $r_s=4$, where
  $a_{\rm B}$ is the Bohr radius.

In this work, we compare the new PIMC simulation data \cite{Filinov23} with  rigorous virial expansions to show\\ (i) the quality of the simulation data,\\ (ii) discuss the question whether higher order virial coefficients can be obtained from the PIMC simulations. \\
We compare with the chemical picture by considering the medium dependent ionization potential and the degree of ionization.
Similar considerations were already performed for the uniform electron gas, for which also high-accurate PIMC simulations were compared with virial expansions \cite{Ropke24}.  Here, the method of virial plots has proven effective to show the virial coefficients clearly. 

The range of temperature of the new PIMC simulations for hydrogen is very interesting. Since temperatures of the order of 150\,000 K are comparable with the binding energy of the hydrogen atom, we have neither the fully ionized plasma nor the atomic gas, but the transition region of a partially ionized plasma. A special feature of the hydrogen system is the formation of bound states, but not only the well-known H atom including all excited states, but also the formation of H$_2$ molecules, and other excited bound states.
 However, in this work, we will not discuss the contribution of other bound states in addition to the H atom. 

In the following we consider a  charge-neutral plasma, with temperature  $T$, total electron density $n_e$, total proton density $n_p$, and $n_p=n_e=n$.  
In a simplified picture, some of the electrons are bound to atoms (H, density $n_{\rm H}$), while the remaining electrons are free (density $n^*_e$).
 The total electron density is decomposed as $n_e=n_{\rm H}+n^*_e$. 
The density of the H$^+$ ions (free protons $p$) is $n^*_p=n^*_e$, due to charge neutrality.
 Since the formation of bound states is a quantum phenomenon, classical descriptions of the hydrogen plasma are only of limited value.
 In an improved picture, correlations in the continuum of the free (scattering) states must be addressed, which is a subject of this paper.
 
In the low-density limit, the concept of the ionization degree $\alpha=n^*_e/n_e$ 
gives the correct limit for the equations of state (EoS), and
the   composition of the plasma is given by the Saha equation.
This chemical picture of the ionization degree becomes more complex if dense plasmas are considered. 
The ionization potential depression (IPD) has been introduced to describe the influence of the medium on the ionisation equilibrium.
At very high densities, the bound states are dissolved, and we observe the transition of the partially ionized plasma to a degenerate $e-p$ liquid. 
The dissolution of the H atoms, denoted as Mott effect, is discussed as a consequence of screening and/or Pauli blocking. 
The bound states become broadened so that their disappearance is not  sudden  (Inglis-Teller effect). Instead, a resonance in the continuum has also some properties similar to  a weakly bound state. 
Therefore, the subdivision of the electrons into   "bound" and "free" electrons is not unique. 
A quantum statistical approach to study these effects is given, for instance in Ref. \cite{KKER86}.

Basic concepts of the quantum statistical approach - spectral functions, self-energy, virial expansions - are considered in Sect. \ref{sec:QuStat}. 
Approximations are discussed in Sect. \ref{sec:Medium}, where the Saha equation and in-medium effects are shown.
The discussion of the PIMC simulations \cite{Filinov23} is performed in Sect. \ref{sec:PIMC}. 
A discussion of the extension of the Saha equation using the quasiparticle concept,  IPD, and ionization degree is found in Sect. \ref{sec:discussion}.
Conclusions are drawn in Sect. \ref{sec:Concl}.

\section{Quantum statistics approach to the equation of state}
\label{sec:QuStat}

We describe the hydrogen plasma as a non-relativistic Coulomb system consisting of electrons and protons. Due to charge neutrality,   the total particle densities of electrons $n_{\rm e}$ and protons $n_{\rm p}$ are equal,  $n=n_{\rm e}=n_{\rm p}$.
We use atomic units $T_{\rm Ha}$ for the temperature $T$ and $n_{\rm Bohr} = n\,a^3_{\rm B}=3/(4 \pi r_s^3)$ for particle densities, 
so that
\begin{equation}
\label{units}
    T_{\rm Ha}=\frac{1}{315777.1}\, \frac{T}{{\rm K}}, \qquad n_{\rm Bohr}= 1.4818471\times 10^{-25} n_{\rm e}\,{\rm cm}^3\,.
\end{equation}

Hydrogen plasma consists of two  components $i$, the electrons $i=$ e and the protons $i=$ p. Equilibrium thermodynamics relates various macroscopic parameters through equations of state. 
For instance, within the grand canonical ensemble, the equations of state 
\begin{equation} \label{density}
n_i(T,\mu_j)=\frac{1}{V}{\rm Tr}\{\rho \hat N_i\}=\frac{1}{V}{\rm Tr}\left\{\rho \sum_k \hat n_{i,k}\right\}, \qquad \rho=\exp\left(\beta \hat H-\sum_i \beta \mu_i \hat N_i\right)/{\rm Tr}\exp\left(\beta \hat H-\sum_i\beta \mu_i \hat N_i\right)
\end{equation}
give the  densities $n_i$ as function of $T$, $\beta=1/k_BT$, and the chemical potentials $\mu_i$.
We use second quantization where single-particle states $|k\rangle = |{\bf k},\sigma \rangle$, given by wave number vector $\bf k$ and spin $\sigma$. 
The occupation number operator $\hat n_{i,k}=a^+_{i,k}a^{}_{i,k}$ is given by the creation and annihilation operators in the corresponding states. 

A fundamental quantity to describe physical properties of a many-particle system is the spectral function, which is also well defined in a dense system.   The Fourier transform of the average of creating an electron at time $t$ in the state $|k\rangle$ and extracting the electron at time $t'$ in the state $|k'\rangle$ (Heisenberg picture)
\begin{equation}
\int d(t'-t) e^{i\omega (t'-t)} \langle a^+_{e,k}(t) a^{}_{e,k'}(t')\rangle =\frac{1}{e^{\beta \omega}+1}A_e(k,\omega ) \delta_{k,k'}
\end{equation}
defines the spectral function $A_e(k,\omega)$ of the electrons (the variables $T, \mu_i$ are not shown). Because of homogeneity in space and time (equilibrium) it is diagonal in momentum representation $k$ and dependent only on $(t'-t)$).  We can rewrite Eq. (\ref{density}) for the electrons as
\begin{equation}
n_e(T, \mu_i) =\frac{1}{V}\sum_k\int \frac{d \omega}{2 \pi} \frac{1}{e^{\beta \omega}+1}A_e(k,\omega ) =\int \frac{d \omega}{2 \pi}  \frac{1}{e^{\beta \omega}+1}D_e(\omega ), 
\end{equation}
where 
\begin{equation}
    D_e(\omega )=\frac{1}{V}\sum_k A_e(k,\omega) =\sum_\sigma\int \frac{d^3 k}{(2 \pi)^3} A_e(k,\omega)
\end{equation}
denotes the electron density of states.
Similar expressions can be obtained for the protons which, however, will be considered as classical particles at the conditions discussed in this work.
%

Having the EoS $n(T,\mu_i)$ to our disposal, other thermodynamic functions can be derived. 
 An inversion, i.e. the solution of this set of equations for the chemical potentials, gives us the relations $\mu_i(T,n_j)$ so that temperature and densities are the independent variables (canonical ensemble).
 In particular, the thermodynamic potential free energy density $f(T,n_i)$ 
is obtained by integration 
\begin{equation}
\label{fp(n)}
f(T,n_i)=\int_0^{n_j} \mu_j(T,n'_j,n_i) dn'_j|_{n_i,i \neq j}, 
\end{equation}
and all  other thermodynamic variables such as pressure are obtained from this quantity.

\subsection{The GF approach and the generalized Beth-Uhlenbeck equation \label{GFandGBUE}}

The method of thermodynamic Green functions delivers a systematic first principle approach to the spectral function and subsequently to the thermodynamic and transport properties of the many-particle system. 
The method of Feynman diagrams allows to describe many-particle effects performing systematic perturbation expansion including partial summations.

The single-particle Green function $G_i(k,z)$ is related to the self-energy $\Sigma_i(k,z)$,
\begin{equation}
G_{i}(k,z)=\frac{1}{E_{i,k}+\Sigma_{i}(k,z)-\mu_{i}-z},\qquad E_{{i},k}=\frac{\hbar^2 k^2}{2 m_{i}}, 
\end{equation}
and to the spectral function
\begin{equation}
A_i(k,\omega) = 2 \lim_{\epsilon \to 0} {\rm Im} G_i(k,\omega-i \epsilon)=2 \frac{{\rm Im} \Sigma_i(k,\omega)}{[E_{i,k}+{\rm Re}\Sigma_i(k,\omega)-\mu_i-z]^2+[{\rm Im} \Sigma_i(k,\omega)]^2}.
\end{equation}
We focus on the electron spectral function and drop the index $i$. For small ${\rm Im} \Sigma(k,\omega)$ we have the expansion
\begin{equation}
\label{Aexpans}
A(k,\omega) \approx\frac{2 \pi \delta(\omega+\mu_e-E_k^{\rm qu})}{1-\frac{d}{dz}{\rm Re}\Sigma(k,z)|_{z=E^{\rm qu}_k-\mu_e}}+2{\rm Im} \Sigma(k,\omega+i0) \frac{d}{d \omega}\frac{{\cal P}}{\omega+\mu_e-E_k^{\rm qu}}.
\end{equation}
The first term is the quasiparticle peak, a sharp $\delta$-like peak at the quasiparticle energy
\begin{equation}
\label{quEnGl}
E_k^{\rm qu}= E_k+{\rm Re}\Sigma(k,\omega)|_{\omega = E^{\rm qu}_k-\mu_e}.
\end{equation}
Correlations, in particular bound states, are obtained from the second part of 
Eq.\,(\ref{Aexpans}). 
For this, a cluster expansion of the self-energy can be performed.
Ladder-T matrices in $\Sigma(k,iz_\nu)$, defined at the Matsubara frequencies $z_\nu= \pi \beta (2 \nu +1)$, are evaluated solving the in-medium $m$-particle Schr{\"o}dinger  equation, derived in \cite{RKKKZ78,ZKKKR78},
\begin{equation}
\label{mSGl}
(E_1^{\rm qu}+\dots E_m^{\rm qu}) \psi_n(1\dots m)+ \sum_{i<j} V^{\rm eff}_{ij,i'j'}[1-f(i)-f(j)]\psi_n(1\dots i' \dots j' \dots m)=E^{\rm qu}_n \psi_n(1\dots m),
\end{equation}
$f(i)=[\exp{(E_{i,k}-\mu_i)/k_BT}+1]^{-1}$ is the Fermi distribution function. 
 This  equation (\ref{mSGl}) contains the screened potential as approximation for $V^{\rm eff}_{ij,i'j'}$ and the Pauli blocking term describing the exchange term due to antisymmetrization. In the approximation considered in \cite{RKKKZ78,ZKKKR78},  
the self-energy contains the Debye shift and the Fock shift (exchange term).

For the electron-proton interaction, the protons are considered as classical in the region considered here so that the Fermi distribution in the Pauli blocking term can be dropped. However, the interaction takes place with all protons so that the structure factor enters. Dynamical screening is not only by free but also by bound electrons, beyond RPA the structure factor is important. Similarly, phase space occupation is not only by free electrons but also by bound electrons.
A systematic treatment of these contributions is given by the Green's function method, see \cite{KKER86}.


In the low-density limit, the standard solution for the H atom is obtained, if 
\begin{equation}    
V^{\rm Coul}(k_1k_2,k'_1k'_2)=\frac{e^2}{\epsilon_0 (k'_1-k_1)^2}\delta({\bf k}_1+{\bf k}_2,{\bf k}'_1+{\bf k}'_2) \delta_{\sigma_1,\sigma_1'}\delta_{\sigma_2,\sigma_2'}
\end{equation}
is taken. Since the Coulomb potential leads to divergencies due to the long-range interaction, we consider  the Debye potential
\begin{equation} 
\label{Veps}
V^{\epsilon}(k_1k_2,k'_1k'_2)=\frac{e^2}{ \epsilon_0 [(k'_1-k_1)^2+\epsilon^2]}\delta({\bf k}_1+{\bf k}_2,{\bf k}'_1+{\bf k}'_2) \delta_{\sigma_1,\sigma_1'}\delta_{\sigma_2,\sigma_2'},
\end{equation}
for which we have $\lim_{\epsilon \to 0} V^{\epsilon}=V^{\rm Coul}$.

In the simple case of the $e - p$ system with (finite range) interaction $V^{\epsilon}$ the isolated two-particle bound state spectrum is labeled by $s,l,m$, ($n$ is already used for the density), atoms are classical  (non-degenerate) and the scattering states are described by the phase shift $\delta_l(E)$. Then we obtain the Beth-Uhlenbeck formula \cite{BU} (the contributions of the $e-e$ channel and $p-p$ channel are analogous, but are not shown here)
\begin{equation}
\label{simpleBU}
n_e(T,\mu)=\frac{2}{V}\sum_k f_e(E_k)+\frac{4}{\Lambda_{\rm H}^3} e^{\beta (\mu_e+\mu_p)} \sum_l (2l+1)\left[\sum_s^{N_l}e^{-\beta E_{sl}}+ \int \frac{dE}{\pi} e^{-\beta E} \frac{d}{dE}  \delta_l(E)\right].
\end{equation}
where $N_l$ is the number of bound states in channel $l$,  the factors 2 and 4, resp., arise from the spin summation, and $\Lambda_i^2=2 \pi \hbar^2/(m_i k_BT)$. The second term of this expression can be written as $4 \Lambda_{\rm H}^{-3} \exp[\beta (\mu_e+\mu_p)] \sigma (T)$ where $\sigma (T)$ is the intrinsic partition function (without the spin factors).
After partial integration and using the Levinson theorem  $\delta_l(0)=N_l \pi$, it reads
\begin{equation}
\label{partIBU}
n_e(T,\mu)=\frac{2}{V}\sum_k f_e(E_k)+\frac{4}{\Lambda_{\rm H}^3} e^{\beta (\mu_e+\mu_p)} \sum_l (2l+1)\left[\sum_s^{N_l}(e^{-\beta E_{sl}}-1)+ \beta \int \frac{dE}{\pi} e^{-\beta E} \delta_l(E)\right].
\end{equation}
Both expressions (\ref{simpleBU}) and (\ref{partIBU}) give rigorous results for the second virial coefficient.
We see that the subdivision of the second virial coefficient into a contribution from bound states and from scattering states is not unique. Integration by parts shifts part of the binding in eq. (\ref{simpleBU}) into the part of scattering states in eq. (\ref{partIBU}).\\

At finite density, the free particle peak is shifted to the quasiparticle energy so that a more consistent approach should be performed with quasiparticle states.
Considering the quasiparticle shifts in mean-field (Hartree-Fock) approximation, we obtain the Generalized Beth-Uhlenbeck formula \cite{Schmidt90} 
\begin{equation} 
\label{BUmedium}
n_e(T,\mu)=\frac{2}{V}\sum_k f_e(E^{\rm qu}_k)+\frac{4}{\Lambda_{\rm H}^3} e^{\beta (\mu_e+\mu_p)} \sum_l (2l+1)\left[\sum_s^{N_l}(e^{-\beta E^{\rm qu}_{sl}}-1)+ \frac{\beta}{\pi}\int_{E_c}^\infty dE e^{-\beta E} \left( \delta^{\rm qu}_l(E)-\frac{1}{2} \sin[2\delta^{\rm qu}_l(E)]\right)\right].
\end{equation}
$E_c$ denotes the continuum edge. In order to avoid double counting if we introduce the quasiparticle energy, we had to extract from the second virial contribution the interaction contribution to $E^{\rm qu}_k$. 
This  leads to the appearance of the sin term. The Hartree-Fock shift is the contribution of the lowest order of interaction and therefore
should not appear in the second part which we denote as the correlated part to the electron density.
This must also be taken into account if improved expressions for $E^{\rm qu}_k$ are derived.
Note that the quasiparticle picture works also in the high-density limit.
To extend the range of applicability of the Beth-Uhlenbeck formula (\ref{partIBU}), we can take also the in-medium results $E_{sl}^{\rm qu}, \delta_l^{\rm qu}(E)$ of the in-medium Schr\"odinger equation (\ref{mSGl}).
The compensation of contributions of the interaction which are already taken into account in the single quasiparticle term must be performed.
In principle, we can also use empirical values for the quasiparticle shifts as known, for example, in nuclear physics.

Eq. (\ref{BUmedium}) gives a simple subdivision of the density: the contribution of the  single quasi-particles
and the correlated part. 
Neglecting the in-medium corrections and the contributions of the continuum, different forms of the Saha equation (mas action law) are obtained.
The correlated part contains the contribution of the bound states and the continuum part, where there is no clear distinction between both parts. 
In Eqs. (\ref{partIBU}), (\ref{BUmedium}), the subtraction of 1 avoids unphysical jumps if a bound state disappears. Since only the total correlated part, the sum of bound and continuum contribution, is well defined, the concept of "bound state density" is not well defined for dense plasmas. 

\subsection{The standard virial expansion and its limits}

We consider first a gas of atoms at temperature $T$ and density $n$ with short-range interaction $V^{\rm sr}(r)$.
The equations of state for other thermodynamic variables are functions of $T,n$. 
We assume an expansion in powers of $n$ such as for the pressure
\begin{equation}
\label{virial}
    p(T,n)=p_1(T) n+p_2(T)n^2+p_3(T)n^3+\dots,
\end{equation} 
$p_1(T) = k_BT$ is the contribution of the ideal gas.  $p_2(T)$ is the second virial coefficient.
From classical statistics the expression 
\begin{equation}
 p_2(T)=k_BT\int d^3r \left[e^{-\beta V^{\rm sr}(r)}-1\right]   
\end{equation} 
is known. 
Virial expansions similar to (\ref{virial}) can be given for other EoS such as the free energy density $f(T,n)$ or the chemical potential $\mu(T,n)$.

The convergence of the series expansion (\ref{virial}) is problematic. For attractive interactions, higher order virial coefficients increase with decreasing temperature  since $-V^{\rm sr}(r)/T$ becomes large. 
For attractive interactions, bound states can occur, and we have to apply quantum physics.
The exact expression for the second virial coefficient has been found by Beth and Uhlenbeck \cite{BU}.
Using the relations (\ref{fp(n)}), 
the Beth-Uhlenbeck formula (\ref{simpleBU}) can be transformed into an expression for the free energy. 
From this, the pressure is obtained in the virial form (\ref{virial}) so that an analytical expression for the second virial coefficient is found.
The sum over the relative motion of the two particles, the intrinsic partition function, has contributions from bound states and the continuum of scattering states.
In general, for a consistent approach, the virial coefficients should contain not only the bound cluster in the ground state but also all excited bound states of this cluster, as well as the contribution of the continuum correlations.

 A serious problem of convergence appears for long-range interactions such as the Coulomb interaction.
At large distance $r$, we can expand the classical second virial coefficient so that $p_2(T)\approx \int_0^\infty 4 \pi r^2 dr [-V(r)]$. 
For $V(r) \propto 1/r$ at large $r$, this integral is not convergent. 
As a consequence, the virial expansion (\ref{virial}) has divergent viral coefficients.
The solution of this mathematical problem (the pressure is not diverging) is that the analytical behavior of $p(T,n)$ is more complex near $n=0$.
As shown in the following Sect. \ref{sec:Coulomb}, other terms containing $n^{1/2}$ appear.
The Coulomb interaction is replaced by the screened potential which is short range, and another series expansion is obtained which contains powers of $n^{1/2}$.

\subsection{Coulomb systems}
\label{sec:Coulomb}

We have seen that owing to the long-range character of the Coulomb interaction, divergences appear in the standard virial expansion (\ref{virial}) in powers of $n$ if the hydrogen plasma is considered. 
At the same time, the Beth-Uhlenbeck formula (\ref{simpleBU}) is no longer applicable since the scattering phase shifts $\delta_l(E)$ cannot be defined in the standard form. 
To circumvent this problem, we can work with the infinitesimal modification of the Coulomb potential (\ref{Veps}), 
but the Green's function method allows another possibility performing partial summations of special diagrams of the perturbation expansion \cite{KKER86}.
The  Hartree term for each component is divergent but cancels to zero because of charge neutrality $n=n_e=n_p$. 
The next contribution to the second virial coefficient which is of second order in the interaction is also divergent. 
This can be cured by partial summation of ring diagrams so that the Debye term appears which is of the order $n^{1/2}$.
A related problem is the intrinsic partition function of the bound states, which also appears in the Beth-Uhlenbeck formula (\ref{simpleBU}) (without spin factors)  
\begin{equation}
    \sigma^0(T)=\sum_l (2l+1) \sum_s  e^{-\beta E_{sl}}
\end{equation}
with $E_{sl}=-1/(2 s^2)$ Ha, which is divergent.
It can be replaced by the Planck-Brillouin-Larkin partition function \cite{KKER86}
\begin{equation} \label{PBL0}
    \sigma^{\rm PBL}(T)=\sum_{slm}[e^{-\beta E_{sl}}-1+\beta E_{sl}]
\end{equation}
which is convergent.
The subtraction -1 has been discussed already in connection with the Beth-Uhlenbeck formula (\ref{partIBU}). 
Similarly, it can be shown that the term $+\beta E_{sl}$ compensates divergent terms which occur in the contribution of the continuum in lowest order of perturbation theory, see \cite{KKER86}. We see that the standard form of the virial expansion (\ref{virial}) is not possible for Coulomb systems, and we have to perform some modifications.

From quantum statistics,  using the method of thermodynamic Green's functions to calculate $n_i(T,\mu_i)$ (\ref{BUmedium}), the virial expansion
for the free energy density (\ref{fp(n)}) 
of the hydrogen plasma as a function of $T$ and the density $n$  is obtained which reads 
\begin{eqnarray}
\label{Fvir}
&&\beta f(T,n)=2 n \ln n + [3/2\ln(2 \pi \hbar^2/(m_e k_BT))-1] n+ [3/2\ln(2 \pi \hbar^2/(m_p k_BT))-1] n 
\nonumber \\ &&\left.-F_{\rm Debye}(T)n^{3/2}-F_1(T)n^2 \ln n-F_2(T) n^2
-F_3(T)n^{5/2} \ln n-F_4(T) n^{5/2}+{\cal O}(n^3\ln n)\right\},
\end{eqnarray}
see  \cite{KKER86} where expressions for the lowest virial coefficients $F_i$ up to $F_3(T)$ are also given. 
Because of the long-range character of the Coulomb interaction, half-number exponents of the density and logarithmic terms appear. 

The virial expansion of the pressure $p(n,T)$ of the hydrogen plasma can be obtained from the free energy density 
\begin{equation}
    p=n \frac{\partial f}{\partial n}-f.
\end{equation}
It can be given in the following form using atomic units, Eq. (\ref{units})
\begin{eqnarray}
\label{virialexp}
\frac{\beta p}{2 n}&=&A_{\rm ideal}(T)-A_{\rm Debye}(T)n_{\rm Bohr}^{1/2}-A_1(T)n_{\rm Bohr} \ln(n_{\rm Bohr}) \nonumber \\
&&-A_2(T)n_{\rm Bohr}-A_3(T)n_{\rm Bohr}^{3/2} \ln(n_{\rm Bohr})-A_4(T) n_{\rm Bohr}^{3/2}+{\cal O}(n_{\rm Bohr} \ln(n_{\rm Bohr})).
\end{eqnarray}
and the  exact expression \cite{Ebeling69}
\begin{eqnarray}
\label{virialp}
A_{\rm ideal}(T)&=&1,\nonumber\\
A_{\rm Debye}(T)&=& \frac{(2 \pi)^{1/2}}{3 T_{\rm Ha}^{3/2}},\nonumber\\
 A_1(T)&=& 0,\nonumber\\
 A_2(T)&=&\frac{ \pi}{T_{\rm Ha}^{3/2}} \left\{\left[Q(\xi_{ee})-\frac{1}{2} E(\xi_{ee})\right]
 +2 \left(\frac{1}{2}+\frac{m_e}{2m_p}\right)^{3/2}Q(\xi_{ep})+\left(\frac{m_e}{m_p}\right)^{3/2}\left[Q(\xi_{pp})-\frac{1}{2} E(\xi_{pp})\right]
\right\}\nonumber \\
&&+\frac{\pi}{12\,T_{\rm Ha}^3} \ln \left[\frac{4 m_e m_p}{(m_e+m_p)^2}\right]
\label{A2(T)}\nonumber\\
&=& A_{2,ee}(T)+ A_{2,ep}(T)+ A_{2,pp}(T)+ A_{2}^{(0)}(T),\nonumber \\
A_3(T)&=&\frac{(2 \pi^3)^{1/2}}{T_{\rm Ha}^{9/2}}
\end{eqnarray}
with 
\begin{eqnarray}
 \xi_{ee}&=&-T_{\rm Ha}^{-1/2},   \nonumber \\
 \xi_{ep}&=&T_{\rm Ha}^{-1/2}(2m_em_p/(m_e+m_p))^{1/2}=T_{\rm Ha}^{-1/2}[2/(1+\gamma)]^{1/2}, \nonumber \\
 \xi_{pp}&=&-T_{\rm Ha}^{-1/2}m_p^{1/2}=-T_{\rm Ha}^{-1/2}/\gamma^{1/2},
\end{eqnarray}    
and $\gamma=m_e/m_p=5.446170226 \times 10^{-4}$.
The second virial coefficient contains the direct ($Q(x)$) and exchange ($E(x)$) functions 
\begin{eqnarray}
\label{Q}
Q(x)&=&-\frac{1}{6}x-\frac{\pi^{1/2}}{8}x^2-\frac{1}{6}\left(\frac{1}{2} C+\ln(3)-\frac{1}{2}\right)x^3
+\sqrt{\pi} \sum_{m=4}^\infty\frac{\zeta(m-2)x^m}{2^m \Gamma(m/2+1) },
\end{eqnarray}
\begin{eqnarray}
 E(x)&=&\frac{\pi^{1/2}}{4}+\frac{1}{2}x+\frac{\pi^{1/2}}{4} \ln(2) x^2+\sqrt{\pi} \sum_{m=3}^\infty\frac{\zeta(m-1)x^m}{2^m \Gamma(m/2+1)}\left(1-\frac{4}{2^m}\right),
\end{eqnarray}
$\zeta(m)$ denotes the Riemann zeta function, and $C=0.5772156649\dots$ is Euler's constant. 
We consider charge-neutral plasmas, the linear term in $Q$ is zero. To have convergent results, a large number of terms in the sum must be considered. 
In particular, this refers to the electron-proton channel $A_{2,ep}(T)$ where $x=\xi_{ep}>0$,  for which bound states occur and a perturbation expansion does not work.  An asymptotic expansion in a semi-convergent series is useful, in particular to calculate the contribution of the $e-p$ part, 
\begin{equation}
\label{Qsigma}
Q(x)=2\pi^{1/2} \Theta(x) \left[\tilde \sigma^{\rm PBL}(x)-\frac{x^2}{8}\right]-\frac{1}{6}x^3\left[\ln|x|+2 C+\ln(3)-\frac{11}{6}\right]-\frac{1}{12}x-\frac{1}{60\,x}+{\cal O}(x^{-3})
\end{equation}
where $\Theta(x)=1$ if $x>0$, and 0 else. The Planck-Brillouin-Larkin partition function (\ref{PBL0}) is then given by (we use another variable $x$)
\begin{equation}
\label{PBL}
\tilde \sigma^{\rm PBL}(x) = \sum_{s=1}^\infty s^2\left(e^{x^2/(4 s^2)}-1- \frac{x^2}{4 s^2}\right) . 
\end{equation}

Results for the second virial coefficient $A_2(T)$ are shown in Tab.\,\ref{Tab:A2} for temperatures  taken from the PIMC simulations \cite{Filinov23}.  
The contributions of the different channels are also shown. 
The $e - e$ - term $A_{2,ee}(T)= \pi/T_{\rm Ha}^{3/2}\left[Q(\xi_{ee})-\frac{1}{2} E(\xi_{ee})\right]$
and the $p - p$ - term $A_{2,pp}(T)= \gamma^{3/2} \pi/T_{\rm Ha}^{3/2}\left[Q(\xi_{pp})-\frac{1}{2} E(\xi_{pp})\right]$ 
describe repulsive interactions ($x < 0$) so that bound states do not occur, we have a continuum of scattering states.
The $e - p$ - term $A_{2,ep}(T)= (1/2+\gamma/2)^{3/2} 2 \pi/T_{\rm Ha}^{3/2}Q(\xi_{ep})$ describes the attractive Coulomb interaction ($x > 0$) where bound states occur.
We show the bound state contribution $A^{\rm bound}_{2,ep}(T)=(1/2+\gamma/2)^{3/2} 4\pi^{3/2}/T_{\rm Ha}^{3/2} \tilde \sigma^{\rm PBL}(T_{\rm Ha}^{-1/2})$ with
(\ref{PBL}) separately.
The remaining part $A^{\rm scat}_{2,ep}(T)$ is the contribution of the scattering states.
It is clearly seen that as the temperature decreases, the bound state component dominates the second virial coefficient,
while the contributions from the continuum of scattering states become negligibly small.

We used the expressions found in Ref. \cite{KKER86}. Note that we introduced the virial coefficients (\ref{virialexp}) dividing $\beta p$ by $2n$ so that $A_{\rm ideal}=1$, so that also the expression for $A_2$ given there has to be divided by 2. 
Tables for the functions $Q(x), E(x)$ are found, e.g., in Ref. \cite{EKK76}.

\begin{table}
    \begin{center}
        \begin{tabular}{|c|c|c|c|c|c|c|c|c|c|c|c|}
            \hline
             $T$[K]& $T_{\rm Ha}$&$\xi_{ee}$ &$-A_{2,ee}$&$\xi_{ep}$ &$A_{2,ep}$& $A^{\rm bound}_{2,ep}$ & $A^{\rm scat}_{2,ep}$ & $\xi_{pp}$ & $A_{2,pp}$ & $A_2^{(0)}$& $A_2$ \\
            \hline
250000 & 0.791702 & -1.12388 & 0.75034 & 1.58897 & -1.58845 & 4.25724 & -5.84495 & -48.1586 & 4.53126 & -3.23415 & 0.459 \\ 
181823 & 0.575798 & -1.31785 & 2.3986 & 1.86321 & -0.562557 & 13.8153 & -14.379 & -56.4702 & 12.2151 & -8.40685 & 5.64428 \\
125000 & 0.395851 & -1.5894 & 8.74376 & 2.24714 & 14.9588 & 57.4093 & -42.4667 & -68.1065 & 39.1746 & -25.8732 & 37.004 \\
62500 & 0.197925 & -2.24776 & 89.233 & 3.17794 & 676.366 & 1008.75 & -332.656 & -96.3171 & 336.797 & -206.986 & 895.411 \\
31250 & 0.0989627 & -3.17881 & 877.899 & 4.49429 & 37892.5 & 40632.7 & -2743.41 & -136.213 & 2881.6 & -1655.88 & 39996.1 \\
15625 & 0.0494814 & -4.49551 & 8417.28 & 6.35588 & 1.743$\times 10^7$ & 1.745$\times 10^7$ & -23336.8 & -192.634 & 24550.6 & -13247.1 & 1.746$\times 10^7$\\
            \hline
        \end{tabular}
    \end{center}
    \caption{Contributions to $A_2(T)$ (\ref{virialp}), for  explicit expressions see (\ref{PBL}) and text below. $A_2^{(0)}(T)=\frac{\pi}{12T_{\rm Ha}^3} \ln \left[\frac{4 \gamma}{(1+\gamma)^2}\right]$ }\label{Tab:A2}
\end{table}

During the last decades,  several papers were devoted to the calculation of the coefficient $A_4(T)$ including  controversial discussions, see Refs. \cite{DeWitt95,AlaPerez,Ebe93,EbFo95,Kahl,AlaMono,Cornu,Riemann}.
However, as outlined below, the study of higher order virial coefficients is beyond the scope of our present work due to the lack of appropriate PIMC data.

\section{Medium modification and the Saha-Debye approach}
\label{sec:Medium}

Historically, another approach has been used to describe the hydrogen plasma at low temperatures where the bound part of the second virial coefficient dominates. In the simplest version, the Saha approach considers the partially ionized plasma as a mixture of free electrons, free protons, and H atoms in the ground state. 
We denote this approximation as Saha0.
In a next step, we include also the excited states of the atom. This leads to the Planck-Brillouin-Larkin partition function, which was already given in the last Section.
The corresponding approximations for the EoS are denoted as $p^{\rm Saha0}(T,n)$, $p^{\rm PBL}(T,n)$, respectively.
Then we discuss the quasiparticle shifts of the free and bound states.  
These quantities can be derived from the generalized Beth-Uhlenbeck formula (\ref{BUmedium})  using corresponding  approximations.
In each case we determine the ionization degree $\alpha=n_e^*/n_e$ with the density of free electrons $n_e^*$.

The use of the Saha approach and improved approximations has recently been discussed in several articles \cite{ebel23,ebel25}. In particular, it has been shown \cite{ebel25} that the Saha approach improves the results of the virial expansion when compared with the results of PIMC simulations.


\subsection{Saha0}
\label{Saha0}
The Saha0 approximation is given as
\begin{equation}
\label{saha0a}
 n_e(T,\mu_i)\approx n^{\rm Saha0}_e(T,\mu_i)=n^{\rm Saha0,*}_e(T,\mu_e)+ n^{\rm Saha0,*}_H(T,\mu_i). 
\end{equation}
The single particle peak of the spectral function is approximated by the free particle peak. The in-medium Schrödinger equation for the hydrogen atom is approximated by the ground state of the free hydrogen atom, 
see also the Beth-Uhlenbeck formula (\ref{simpleBU}) if the excited bound states and the continuum contributions are neglected.
At low-densities, the Fermi distribution function can be replaced by the Boltzmann distribution function.  
We obtain
\begin{equation}
\label{saha0b}
  n^{\rm Saha0,*}_e(T,\mu_e)=\frac{2}{\Lambda_e^3}e^{\beta \mu_e},\qquad \,n^{\rm Saha0,*}_p(T,\mu_p)=\frac{2}{\Lambda_p^3}e^{\beta \mu_p},\qquad\,e^{\beta \mu_p}=\frac{\Lambda_p^3}{\Lambda_e^3}e^{\beta \mu_e}  
\end{equation}
from charge neutrality. With $\alpha^{\rm Saha0}=n_e^{\rm Saha0,*}/n_e^{\rm Saha0}$, 
\begin{equation}
\label{saha0c}
n_H^{\rm Saha0,*}(T,\mu_i)=\frac{4}{\Lambda_H^3}e^{-\beta E_{1s}+\beta \mu_p+\beta \mu_e}  
\end{equation}
and
\begin{equation}
\label{sig0}
    \rho^{(0)}(T)=\frac{\Lambda_H^3}{\Lambda_e^3\Lambda_p^3}\frac{1}{\sigma^{(0)}(T)}, \qquad\, \sigma^{(0)}(T)=2 e^{-\beta E_{1s}},
\end{equation}
and replacing the chemical potentials by the densities with the approximation $n_e\approx n^{\rm Saha0}_e$, we find the solution
\begin{equation}
    \alpha^{\rm Saha0}(T,n_e)=-\frac{\rho^{(0)}(T)}{n_e}+\left(\frac{{\rho^{(0)}}^2(T)}{n_e^2}+2 \frac{\rho^{(0)}(T)}{n_e}\right)^{1/2}.
\end{equation}
For the pressure in this approximation we have
\begin{equation}
\label{pSaha0}
    \frac{\beta}{2 n_e }p^{\rm Saha0}(T,n_e)=
    \frac{1}{2}\left[1+\alpha^{\rm Saha0}(T,n_e)\right].
\end{equation}
This fairly simple approach represents already important properties. The composition is described in good approximation for low temperatures $T_{\rm Ha} \le 1$, see the discussion in Sect. \ref{sec:BoundSaha}.
We have with the general relation
\begin{eqnarray}    
   \label{pTm}
    p(T, \mu_e)&=&\int_{-\infty}^{\mu_e} 2n_e(T,\mu'_e) d \mu'_e  \\
    \beta p^{\rm Saha0}(T,\mu_e)&=&2 n^*_e+n^*_{\rm H}=\frac{4}{\Lambda_e^3} e^{\beta \mu_e}+\frac{4}{\Lambda_e^3}\frac{\Lambda_p^3}{\Lambda_{\rm H}^3} e^{2 \beta \mu_e}e^{-\beta E_{10}}.
\end{eqnarray}
After eliminating $\mu_e$ using the relations (\ref{saha0a}) for the total electron density $n_e$, and $ \Lambda_p^3/\Lambda_{\rm H}^3\approx 1$, we have within the Saha0 model the relation
\begin{equation}
\label{ionpa}
    e^{-\beta E_{10}}=\frac{1}{2n \Lambda_e^3}\,\frac{1-p^{\rm Saha0}/(2 nT)}{(0.5-p^{\rm Saha0}/(2 nT))^2}.
\end{equation}
We use this relation to define the effective ionization potential in Saha0 approximation (in Hartree units)
\begin{equation}
\label{eq:Ieffec0}
    I^{\rm Saha0}_{\rm Ha}(T,n)=T_{\rm Ha}\ln\left[\left( \frac{T_{\rm Ha}}{2 \pi}\right)^{3/2} \frac{1}{2 n_{\rm Bohr}}\frac{1-p/(2 nT)}{(0.5-p/(2 nT))^2}\right].
\end{equation}
In the region, where the Saha0 model is a good approximation, we have with (\ref{ionpa}) the relation
\begin{equation}
\label{ionp}
    I^{\rm Saha0}_{\rm Ha}(T,n)\approx \frac{1}{2}.
\end{equation}
The effective ionization potential is the binding energy 0.5 Ha of the hydrogen ground state.

\subsection{The Planck-Brillouin-Larkin partition function}

It is straight forward to extend the "Saha0" approximation by including excited states of the hydrogen atom, energies $E_{sl}$.
Instead of $\sigma^{(0)}$, eq. (\ref{sig0}), we introduce the intrinsic partition function 
\begin{equation}
    \sigma^{\rm bound}(T)=\sum_{sl}(2l+1)e^{-\beta E_{sl}}.
\end{equation}
As discussed in Sect. \ref{sec:Coulomb}, this expression diverges for the hydrogen atom and must be replaced by the Planck-Brillouin-Larkin partition function (\ref{PBL}), 
\begin{equation}
\label{PBL1}
 \sigma^{\rm PBL}(T) = \sum_{s=1}^\infty s^2\left(e^{-\beta E_s}-1+ \beta E_s\right) . 
\end{equation}
where $E_s =-1/(2 s^2)\, [{\rm Ha}]$. We have to replace in Eqs. (\ref{sig0}) - (\ref{pSaha0}) the quantities $\rho^{(0)}$ by $\rho^{\rm PBL}$, $\alpha^{\rm Saha0}$ by $\alpha^{\rm PBL}$, and $p^{\rm Saha0}$ by $p^{\rm PBL}$ so that
\begin{equation}
\label{pSaha0}
    \frac{\beta}{2 n_e}p^{\rm PBL}(T,n_e)=
    \frac{1}{2}\left[1+\alpha^{\rm PBL}(T,n_e)\right].
\end{equation}

A consistent approximation should not only consider the bound state contribution to the intrinsic partition function of the H atom, but also the contribution to the continuum correlations, see the Beth-Uhlenbeck formula (\ref{partIBU}).
However, the appearance of an additional term $+ \beta E_s$ in Eq. (\ref{PBL1}) indicates that another mathematical problem arises for the Coulomb interaction when applying the Beth-Uhlenbeck formula. 
We cannot define usual scattering phase shifts for the Coulomb potential. 
The virial expansion does not converge.

This problem is related to the analytical form of the virial expansion.
To solve it, the screening of the Coulomb interaction must be considered, which will be described in the following Sect. \ref{sec:mediumSingle}.
Owing to the particularities of the Coulomb potential, the expressions for the second virial coefficient (\ref{Q}), (\ref{Qsigma}) should be used, where terms with powers of $n^{1/2}$ occur.
These terms cannot be obtained from a power series expansion with respect to $n$.

\subsection{Medium modifications of the single-particle states}
\label{sec:mediumSingle}


Within the quantum statistical approach, the low-density form of the spectral function is described taking in-medium effects into account. 
Instead of free particles, quasiparticles are considered, see Eq. (\ref{quEnGl}).
In the lowest order of interaction ($e^2$), the self-energy is given by the Hartree-Fock term
\begin{equation}
\label{eq:HF}
    \Sigma^{\rm HF}(k) = \sum_{k'}[V(k,k';k,k')-V(k,k';k',k)] f(k').
\end{equation}
The Hartree term vanishes due to charge neutrality. 
Expressions for the Fock term are found in Ref. \cite{KKER86}. 
The Fock shift is of the order $n$. 
It is of relevance in regions of the density-temperature plane where the electrons are degenerate.
In our present work, we restrict ourself to the non-degenerate case so that the Fock shift is less important.

For the long-range Coulomb interaction, convergence problems appear with the "direct" contributions to the self-energy (i.e. no exchange terms),
which are of the next order ($e^4$). 
This problem is cured performing a partial summation of ring diagrams, introducing the screened interaction.
In the framework of the Green's function method, the polarization function and the dielectric function $\epsilon(q,\omega)$ are introduced, see Ref. \cite{Reinholz05}.
In random phase approximation (RPA), the polarization function is obtained from the product of free particle propagators.
A detailed discussion is given in Ref. \cite{KKER86}, Sect. 4.3.2. 
In particular, the mean value of the real part of the self-energy gives in the low-density limit the Debye shift
\begin{equation}
\label{Delta0}
    \langle \Sigma_c \rangle=-\frac{e_c^2 \kappa}{8 \pi \epsilon_0}, \qquad \kappa^2=\sum_c\frac{e_c^2n_c}{\epsilon_0k_BT}=\frac{2 e_c^2n}{\epsilon_0k_BT},
\end{equation}
with $c=\{e,p\}$ and $n_e=n_p=n$.
This is a classical effect.  In the low-density limit, this shift in the order of $n^{1/2}$ is stronger compared to other effects which are of the order of $n$. The quasiparticle shift $\Sigma_e(k)$ depends in general on the quantum number $k$, but can replaced by the average shift $\langle \Sigma_c \rangle$ which gives the correct result for the free quasiparticle contribution  in Eq. (\ref{BUmedium}) in first order expansion. The shift of the single quasiparticle energies leads to the lowering $\Delta E_{\rm cont}$ of the continuum of scattering states.

The electrons and ions in the medium should not considered as free particles but show also correlations.
For these improvements, the polarization function must be calculated in approximations beyond the RPA expression.
Part of electrons are bound to atoms and do not participate in the screening.
Improved versions should consider only the screening by free, delocalized electrons and ions (densities $n^*_e=n^*_p=n^*$), not the bound ones.
Also the ions interact and can be strongly correlated, as expressed by the ion structure factor.
Based on the fluctuation-dissipation theorem, the ionic structure factor was considered in the context of the self-energy by 
 Lin et al. \cite{Lin17}.

We shortly outline this approach \cite{Lin17,Lin19}. Using structure factors to characterize the ionic coupling effect and also the electron-ion correlation, the lowering of the continuum  for partially ionized hydrogen plasmas can be expressed as
\begin{equation}
\label{contedge}
	\Delta E_{\rm cont} = \frac{z_c^2 e^2 \kappa^2_\mathrm{eff} a_\mathrm{B}}{8 \pi^2 \epsilon_0 r_c^2} \int_0^\infty \frac{d k}{k} \left[ S_{zz}^\mathrm{ion}(k) +  S_{zz}^\mathrm{el}(k)\right]
\end{equation}
where $a_\mathrm{B}$ is the Bohr radius, $z_c$ denotes the charge number of ion/atom after ionization with $z_c = 1$ for the hydrogen atom, and $r_c = [3 z_c/(4 \pi n_e^*)]^{1/3}$ is the corresponding effective ionic radius. The nonlinear screening parameter $\kappa_\mathrm{eff}$ in Eq.~(\ref{contedge}) reads
\begin{equation}
	\kappa^2_\mathrm{eff} = \frac{3 \Gamma_c}{\sqrt{1 - 0.4 (3 \Gamma_c \alpha_0^2)^{3/4} + 3 \Gamma_c \alpha_0^2}}.
\end{equation}
Here $\alpha_0 = [4/(9 \pi)]^{1/3}$, and the impurity-perturber coupling strength is $\Gamma_c = z_c / (2 \pi \epsilon_0 r_c k_B T)$. Neglecting the influence of dynamical screening effect of free electrons, the electron structure factor can be given by
\begin{equation}
	S_{zz}^\mathrm{el}(k) = \frac{k^2}{k^2 + \kappa_e^{*2}} ,
\end{equation}
with the inverse Debye screening length of free electrons $ \kappa_e^*=[n_e^* e^2/(\epsilon_0k_BT)]^{1/2}$. 
The ionic structure factor is described by the expression
\begin{equation}
	S_{zz}^\mathrm{ion}(k) = [1 - q_\mathrm{scr}(k)]^2 S_\mathrm{pp}(k).
\end{equation}
The function $q_\mathrm{scr}(k) = \epsilon^{-1}_\mathrm{ee}(k)-1$ describes the screening cloud of free electrons around the ion. In the long-wavelength limit, the screening function $q_\mathrm{scr}(k)$ is approximated by $\kappa_e^{*2}/(k^2 + \kappa_e^{*2})$. Generally, the structure factors in Eq.~(\ref{contedge}) have to be calculated using numerical simulations or statistical theory such as the Ornstein–Zernike relation. More details are given in Ref. \cite{Lin17}. In the case of weakly coupled plasmas, the Debye-H\"uckel structure factor for ions $S_\mathrm{pp}(k) = (k^2 + \kappa_e^{*2}) / (k^2 + \kappa_e^{*2} + \kappa_i^{*2})$ can be implemented, with which the Debye-H\"uckel expression [replacing $n$ by $n^*$ in Eq. (\ref{Delta0})] for the lowering of the edge of continuum states $\Delta E_{\rm cont}$ is reproduced,
\begin{eqnarray}
\label{debyeshift}
    \Delta E_{\rm cont}&=&-2 \frac{\kappa^*}{2} \frac{e^2}{4 \pi \epsilon_0} = -2\Delta^{\rm Deb}(T,n),\nonumber \\
    \Delta^{\rm Deb}_{\rm Ha}(T,n)&=&\frac{(2 \pi)^{1/2}}{T_{\rm Ha}^{1/2}}{n^*_{\rm Bohr}}^{1/2}.
\end{eqnarray}

\subsection{Medium modifications of the two-particle states}
\label{sec:mediumTwo}

For short-range interaction, the second virial coefficient is given by the solution of the two-body problem.
From quantum statistics, the Beth-Uhlenbeck formula results. 
For Coulomb systems with long-range interaction, screening must be taken into account.

For the two-particle $e-p$ states, from the Green's function approach the in-medium Schr\"odinger equation is derived \cite{RKKKZ78,ZKKKR78}
\begin{eqnarray}
\label{imSGl}
    && \frac{p^2}{2 m_e}\psi_n(p)+\sum_q V^s(q)\psi_n(p+q)-E_n\psi_n(p)\nonumber \\
    &&=\sum_q V^s(q)\left[\psi_n(p+q) f_e(p)-\psi_n(p)f_e(p+q)\right]
\end{eqnarray}
with the screened potential $V^s(q)$ which is in simplest approximation the Debye potential. 
In general, the dynamically screened Coulomb potential 
\begin{equation}
\label{screenV}
    V^s_{ei}(q, \omega)=V_{ei}(q)\left[1+\int \frac{d \omega'}{\pi}\frac{{\rm Im}\,\epsilon^{-1}(q,\omega'-i 0)}{\omega-\omega'}\right]
\end{equation}
can be introduced, see Ref. \cite{KKER86}.
The influence of the medium is represented by different effects, the screening of the interaction, and the Pauli exchange terms which lead to the right hand side of Eq. (\ref{imSGl}). Both effects describe the weakening of the interaction and are relevant for the dissolution of the bound state at high densities, the Mott effect. 
We discuss both effects separately.

Screening of the Coulomb potential, Eq. (\ref{screenV}), was discussed in connection with the single-particle self-energy.
It is present also in the nondegenerate case.
As shown in Sect. \ref{sec:mediumSingle}, the leading term for the shift of the single-particle states forming the continuum is given by the Debye shift $- \kappa e^2/(8 \pi \epsilon_0) \propto n^{1/2}$.
As shown in \cite{RKKKZ78,ZKKKR78}, this self-energy shift is exactly compensated by the screening of the potential in the order $n^{1/2}$ so that no shift of the bound state energy is obtained in this order of density.
The reason is the charge neutrality of the bound state. Only terms $\propto n$ can occur from the screening of the Coulomb potential. 
The shift and the broadening of the hydrogen bound states in a dense plasma is studied in connection with the line shapes of optical spectra, see, e.g., \cite{Guenter}, where further references can be found.

The right hand side of Eq. (\ref{imSGl}) contains the Fermi function of the electrons as a factor, which is $\propto n$ in the low-density limit.
The last term is the Fock self-energy, if $V^s(q)$ is replaced by the Coulomb potential $V(q)$.
It gives a shift for the scattering (free) states (\ref{eq:HF}), described as $\Delta E^{\rm Fock}(p)$, as well as for the bound states $\Delta E^{\rm bound, Fock}_n$.
The shift of the hydrogen ground state energy owing to this term is for $T=0$ \cite{Eb09}
\begin{equation}
\label{eq:Fockb}
    \Delta E^{\rm bound, Fock}_n=-\frac{n}{2}\sum_{q < q_F}\psi_{1s}^2(q)\frac{e^2}{\epsilon_0 q^2} =- 10\, \pi\, n a_B^3\, [{\rm Ha}]\,.
\end{equation}
The first term on the right hand side of Eq. (\ref{imSGl}) is the Pauli blocking. 
Due to the Pauli principle, states within the Fermi sphere cannot be used to form the bound state, the attractive Coulomb interaction is less efficient. 
We obtain in the low-density limit for the hydrogen ground state energy at $T=0$
\begin{equation}
\label{eq:Pauli}
    \Delta E^{\rm bound, Pauli}_n=\frac{n}{2}\sum_{q < q_F}\psi_{1s}(q)\psi_{1s}(0)\frac{e^2}{\epsilon_0 q^2} =16\, \pi\, n a_B^3\, [{\rm Ha}]\,.
\end{equation}
We see the partial compensation of both effects, both are of the same order with respect to $e^2$ and $n$.
Calculations of the shifts can simply be performed for finite $T$ and a variational ansatz for the ground state wave function.
These shifts are important for degenerate plasmas \cite{Eb09,Lin2}.

With increasing density, another effect becomes important.
The electrons don't interact  independently with the ions, but we have to replace the Coulomb potential of the single proton by the potential of all ions. 
With a second ion, the ground state splits off into two molecular levels.
Then, we have to introduce the formation of H$_2$ molecules.
In general, we have to introduce the ion structure factor as already done for the self-energy \cite{Lin17}.
The use of DFT-MD simulations \cite{Mandy20} is discussed in Sect. \ref{sec:ionization}.
Semi-empirical models to describe the shift of the energy levels due to Pauli blocking are the average atom models where the atomic wave function is confined to a finite volume. 
For various versions of the average atom model see, e.g., Refs. \cite{Yuan02,Yuan24}.

In the present work, formation of bound states other than the H atom are not considered.
Furthermore, the PIMC simulations \cite{Filinov23} to be discussed are performed for nondegenerate plasmas, and the effects of Pauli blocking are small.

\subsection{Saha-virial expansion}
\label{sec:SahaDeb}

Before we discuss the PIMC simulations which, in contrast to DFT-MD simulations, exactly take into account the electron-electron correlations, we summarize the rigorous results obtained from the analytical approach.

The virial expansion is not always appropriate. If bound states are formed, the cluster with the largest binding energy pro particle dominates at low temperatures. 
In this work we restrict us to the hydrogen atom and its excited states, neglecting other bound states such as molecules which become important in special regions of the phase diagram (low temperatures and high densities).
An appropriate starting point is the Saha equation, in the form of Saha0, the so-called chemical picture. 
The plasma is assumed to consists of three components, the free electrons, free protons, and atoms in the ground state. 
Each component is non-interacting with exception of occasional collisions where reactions are possible to sustain the chemical equilibrium.
Degree of ionization and ionization potential are well defined.

This approximation represents a rudimental account of the second virial coefficient. 
This is improved if all excited bound states are taken into account, and we take the second virial coefficient in the PBL approximation.
The contribution of the continuum of scattering states is discarded.
Furthermore, the Fermi or Bose distributions are replaced by the Boltzmann distribution in the non-degenerate case. 

However, this improved chemical picture is not consistent. 
The interaction is partially treated to form bound states, but the constituents are moving freely with exception of occasional reactive collisions which support the chemical equilibrium. We have to consider the interaction between the constituents.
This gives the contribution of the continuum to the second virial coefficient so that the second order of density is exactly reproduced.

An important step is to introduce quasiparticles which take, for instance, mean-field effects into account.
In lowest approximation, the single-particle quasiparticle shift contains the Fock term and the Debye shift.
To be consistent, these contributions of the interaction must be eliminated from the continuum part of the second virial coefficient as described by the generalized Beth-Uhlenbeck formula (\ref{BUmedium}). 
For Coulomb systems, we have the full expression for the second virial coefficient to our disposal.
After quasiparticles are introduced, the remaining part of the second virial coefficient describes the contribution of the continuum in the second order of density.
As shown for the generalized Beth-Uhlenbeck formula (\ref{BUmedium}), the introduction of the quasiparticle shifts gives the expression
\begin{equation} 
\label{SahaDeb}
n_e(T,\mu)=\frac{2}{V}\sum_k e^{-\beta (E^{\rm qu}_k-\mu_e)}+\frac{4}{\Lambda_a^3} e^{\beta (\mu_e+\mu_p-\Delta E_{\rm cont})} \sum_l (2l+1)\left[\sum_s^{N_l}\left(e^{-\beta \tilde E^{\rm qu}_{sl}}-1+\beta \tilde E^{\rm qu}_{sl}\right)+ {\rm cont.~corr.}\right].
\end{equation}
The Planck-Brillouin-Larkin partition function is expressed by the bound state energies $\tilde E^{\rm qu}_{sl}=E^{\rm qu}_{sl}-\Delta E_{\rm cont}$ relative to the edge of the continuum, in Debye approximation $\Delta E_{\rm cont}=2 \Delta^{\rm Deb}$.
The contribution of the continuum correlations (cont. corr.) is often neglected, and in the Saha-Debye0 approximation the contribution of all excited bound states is also neglected. We discuss these approximations in Sect. \ref{sec:BoundSaha}.

Eq. (\ref{SahaDeb}) contains a generalization of the Planck–Brillouin–Larkin partition function  to account for density effects. 
The medium modification of the contribution of the bound states to thermodynamic properties has been extensively discussed \cite{Ebeling85,Zaghloul13}, for instance, by introducing an ionization potential depression dependent cutoff for the maximum principal quantum number, or by directly incorporating such an ionization potential depression dependent cutoff into the Planck–Brillouin–Larkin partition function. 

In the framework of the Saha-Debye0 model, we approximate the single quasiparticle energies as (\ref{debyeshift})
\begin{equation}
\label{ekquasi}
    E_k^{\rm qu}= \frac{\hbar^2 k^2}{2 m_c}- \Delta^{\rm Deb}(T,n), \qquad \Delta_{\rm Ha}^{\rm Deb}(T,n)=\frac{(2 \pi\alpha n_{\rm Bohr})^{1/2}}{T_{\rm Ha}^{1/2}}
\end{equation}
since the Debye shift is performed only by the free particles $n_e^*=\alpha n_e=n_p^*$.
The bound states have no Debye shift so that $E_{10}^{\rm qu}\approx E_{10}$, the shift of the bound state energy is of the order ${\cal O}(n)$. 

The ionization degree must be calculated self-consistently.
The total electron density $n_e=n^*_e+n^*_{\rm H}$ gives the ionization degree $\alpha=n^*_e/n_e$ in Saha-Debye approximation  
\begin{equation}
\label{alphaDeb}
 \alpha+\alpha^2\frac{(2 \pi)^{3/2}n_{\rm Bohr}}{T_{\rm Ha}^{3/2}}\sum_s s^2\left[e^{\frac{1}{2 s^2 T_{\rm Ha}}-\frac{2}{T_{\rm Ha}} \Delta^{\rm Deb}_{\rm Ha}}-1-\frac{1}{2 s^2 T_{\rm Ha}}+\frac{2}{T_{\rm Ha}} \Delta^{\rm Deb}_{\rm Ha}\right] \Theta\left(\frac{1}{2 s^2 T_{\rm Ha}}-\frac{2}{T_{\rm Ha}} \Delta^{\rm Deb}_{\rm Ha}\right)=1,
\end{equation}
with the step function $\Theta(x)=1$ if $x>0$ and =0 else.

Within the Saha-Debye approximation, instead of Eq. (\ref{eq:Ieffec0}) we can also introduce an effective ionisation potential
\begin{equation}
\label{eq:Ieffec}
    I^{\rm Saha-Deb}_{\rm Ha}(T,n)=T_{\rm Ha}\ln\left[\left( \frac{T_{\rm Ha}}{2 \pi}\right)^{3/2} \frac{1}{ n_{\rm Bohr}}\frac{1-\alpha_{\rm Deb}}{\alpha_{\rm Deb}^2}\right],
\end{equation}
where
\begin{equation}
\label{alphaDeb}
    \alpha_{\rm Deb}=\left[\frac{(2 \pi n_{\rm Bohr})^{1/2}}{3 T_{\rm Ha}^{3/2}} +\left( \frac{2 \pi n_{\rm Bohr}}{9T^3_{\rm Ha}}+\frac{\beta p}{n}-1\right)^{1/2}\right]^2\,.
\end{equation}
If we take the expressions for the density and the pressure given in this section, the expression for the effective ionisation potential (\ref{eq:Ieffec}) takes the form 
\begin{equation}
\label{ISahaDeb}
    I_{\rm Ha}^{\rm Saha-Deb}(T,n)\approx 0.5-2\Delta_{\rm Ha}^{\rm Deb}(T,n) \,.
\end{equation}
This low-density result is improved if further effects are taken into account, in particular the improved self-energies discussed in Sects. \ref{sec:mediumSingle}, \ref{sec:mediumTwo}. 
Using the derivation of the pressure according Eq. (\ref{pTm}), the dependence of the quasiparticle shifts on $\mu_e$ must be taken into account.
In the order ${\cal O}(n)$ also other effects appear which are related to the degeneration of the electron system such as the Hartree-Fock terms, the Pauli blocking, and the virial terms connected with the Fermi and Bose functions.

To go to higher densities, the density expansion of the shifts of the single-particle and bound state energies is proposed.
Such medium modifications of the single- and two-particle states are obtained from the Green's functions approaches, see Sects. \ref{sec:mediumSingle} and \ref{sec:mediumTwo}. 
However, the remaining part of the continuum in the higher virial coefficient remains open.
This contribution can be estimated if good results from PIMC simulations are available.

\section{PIMC results of Filinov and Bonitz}
\label{sec:PIMC}

PIMC simulations can be considered as a first-principle approach to evaluate the thermodynamic properties of a many-body system starting from a given Hamiltonian. In particular, correlations are treated avoiding any approximations. The method is described in several publications and will not be repeated here \cite{Dornheim2018,PIMC1,PIMC2}. High-accurate results were obtained for the UEG for a wide range of the thermodynamic parameters temperature $T$ and electron density $n$, see \cite{TD,R24}. Recently, PIMC simulations at extreme conditions, at low densities, were published by Filinov and Bonitz \cite{Filinov23} and shall be discussed in our current work.
Obvious errors in \cite{Filinov23}, Eq. (4) and the pressure value for $r_s=40, T=181823$ K, are not relevant, we also exclude the value $r_s=3.5, T=62500$ K (N64). 
We assume that the main problems with respect to convergence, size-effects, sign problem are under control. 
The sign problems occur if the density becomes high ($r_s < 4$) so that densities relevant for the Mott effect are not in reach.

At high densities, DFT-MD simulations are well established for warm dense matter.
The comparison of PIMC results with DFT-MD simulations has been performed  in Ref. \cite{Filinov23}.
Of interest is the treatment of electron correlations since, in DFT simulations, they are only approximated  choosing the exchange-correlation part of the energy-density functional. On the other hand, the accuracy of the  PIMC simulations is limited by the finite particle number and the number of time slices, caused by computer capabilities.

In this work, we compare the PIMC simulations with benchmarks from the analytical approaches. 
Virial expansions will be applied to check the performance of the simulations.


\subsection{PIMC simulation data}

Systematic PIMC simulations for a lattice of parameter values for $T$ and $n$ is a prerequisite to treat the thermodynamic properties of the plasma and to check simple approximations.
The data for H-plasmas \cite{Filinov23} are stated to have the accuracy of $10^{-4}$. 
However, the region of  free electron density is limited to  $n^* \approx 10^{23}$ cm$^{-3}$. For higher densities, fermionic PIMC simulations were not possible due to the fermion sign problem.
In particular, the interesting region of the Mott transition is not quite reached.
In this work, we are interested in the low-density region where virial expansions can be applied.

Pair distribution functions show the formation of atoms in the $e-p$ pair distribution function, but also the formation of molecules in the $p-p$ pair distribution function. 
In Ref. \cite{Filinov23}, a particular semiempirical approach is proposed to define the amount of atoms/molecules, choosing a distance in the pair distribution function and a time interval, in which the electron remains near the ion.
To discuss the results, the chemical picture was considered with the concepts of molecules, atoms, and free electrons and protons.
Even though the PIMC approach does
not distinguish between bound and free electrons, an (artificial) distinction can be introduced via a cluster analysis.
If two “neutrals” are found at a distance $d_{ii} \le 1.9\, ({\rm or} \,2.25)\,a_B$ , they are counted as a molecule, and the part of free electrons is  given also by a geometrical argument in Ref.  \cite{Filinov23}.
Obviously, these concepts are not well defined, but rather arbitrary, so that they are questionable at increasing density.

In \cite{TD,R23} a method has been presented to extract the virial coefficients from simulated data, provided they have high accuracy. We briefly explain this approach.
The idea is that at fixed $T$, in the low-density limit the lowest virial coefficient will dominate. 
Thus, subtracting the $(m-1)$ lowest virial coefficients from the thermodynamic quantity,
the remaining part allows to determine the next virial term $A_m$ \cite{R23}. 
We focus on the second virial coefficient,
since analytical expressions for $A_2(T)$ are known, see Sect. \ref{sec:Coulomb}. We use the virial expansion for the pressure (\ref{virialexp}) and neglect the virial terms $A_m$ with $m\ge 3$. Thus we can get a virial plot for an effective second viral term  
\begin{equation} \label{A2eff}
A_2^{\rm eff}(T,n_{\rm Bohr})=\frac{1}{n_{\rm Bohr}} \left[-\frac{\beta p}{2n} +A_{\rm ideal}(T)-A_{\rm Debye}(T)n_{\rm Bohr}^{1/2}-A_1(T)n_{\rm Bohr} \ln(n_{\rm Bohr})\right].
\end{equation}
If we assume that the next order virial term $A_3(T)$ gives the leading correction term in the low-density limit, 
a virial plot of $A_2^{\rm eff}(T,n_{\rm Bohr})$ against the abscissa $n_{\rm Bohr}^{1/2} \ln( n_{\rm Bohr})$ gives an effective $A_3^{\rm eff}(T,n_{\rm Bohr})$ as the slope, see Eq.\,(\ref{virialexp}).

\subsection{Benchmarks and accuracy}

\begin{figure}[htp]
\centerline{\includegraphics[width=0.6 \textwidth]{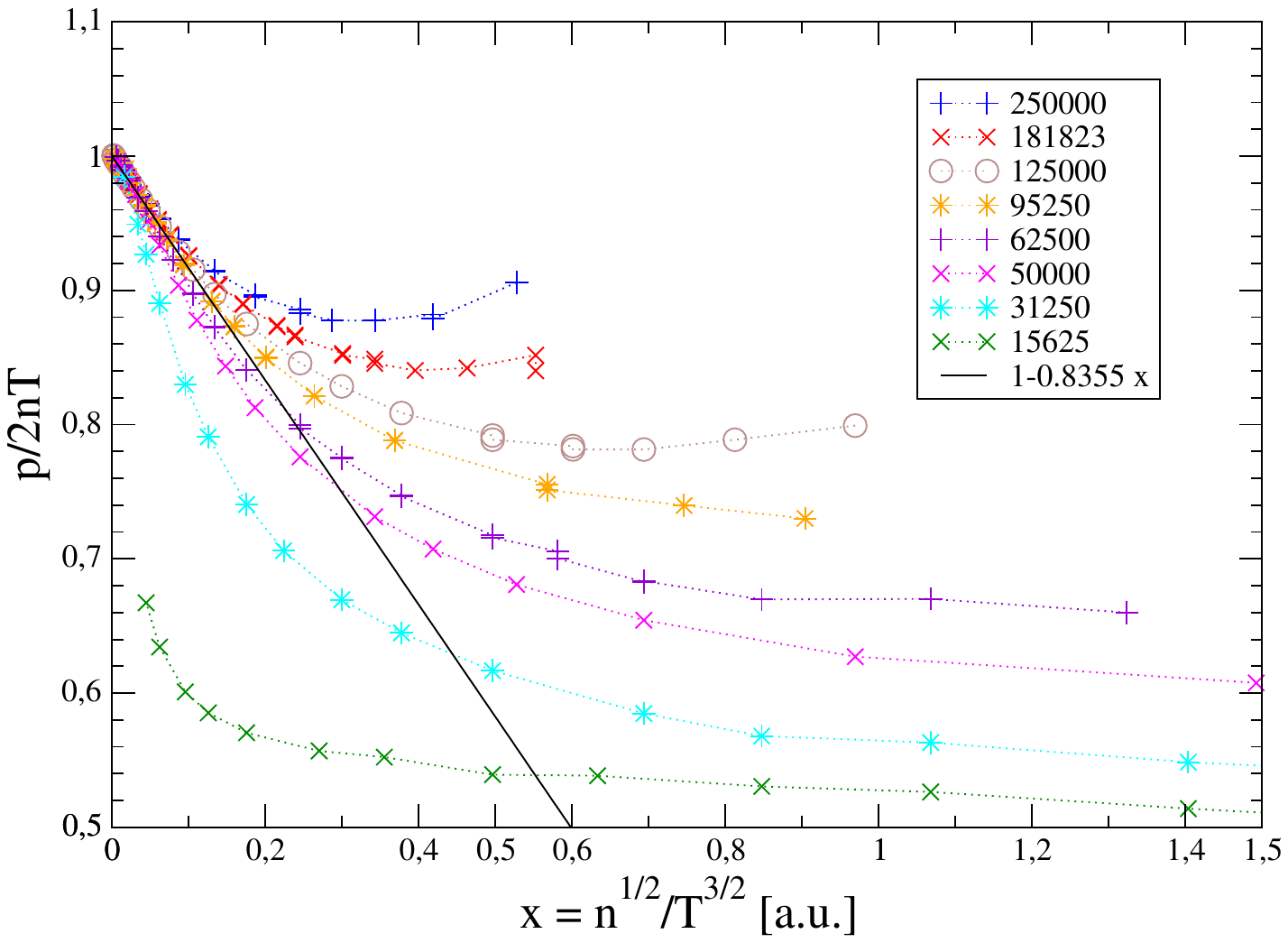} }
\caption{$\beta p/(2 n)$ from PIMC simulations \cite{Filinov23} are shown as a function of $x=n_{\rm Bohr}^{1/2}/T_{\rm Ha}^{3/2}$ for different temperatures in K. The Debye limit $1- (2 \pi)^{1/2} x/3$ is  shown as black line. 
\label{fig:1}}
\end{figure}

We use virial plots to analyze the PIMC data. As a first benchmark we investigate the lowest term of the virial expansion Equ\,(\ref{virialexp}).
We find the expression 
\begin{equation} \label{pdebye}
\frac{\beta p}{2 n}= 
A_{\rm ideal}(T)-A_{\rm Debye}(T)n_{\rm Bohr}^{1/2}+ {\cal O}[n]=
1-\frac{\kappa^3}{48 \pi n}+ {\cal O}[n] =
1-\frac{ (2 \pi n_{\rm Bohr})^{1/2}}{3 T_{\rm Ha}^{3/2}}+ {\cal O}[n] 
\end{equation}
from the theoretical values for the virial coefficients $A_{\rm ideal}$ and $A_{\rm Debye}$, see (\ref{virialp}).
Therefore we choose a virial plot, where we show the data of \cite{Filinov23} for $\beta p/(2 n)$ as function of $n_{\rm Bohr}^{1/2}/T_{\rm Ha}^{3/2}$, see  Fig. \ref{fig:1}. 
We use the units 1 Mbar= $10^{12}$ erg/cm$^3$, $k_{\rm B}=1.380658 (12) \times 10^{-16}$ erg/K, $a_{\rm B}= 5.29177210544 (82) \times 10^{-9}$ cm and calculate
\begin{equation}
\frac{\beta p}{2 n}= \gamma \frac{p/{\rm Mbar}\times r_s^3}{T/{\rm K}}, \qquad \gamma =\frac{2 \pi a_{\rm B}^3 {\rm Mbar}}{3 k_{\rm B} {\rm K} {\rm cm}^3}=2247.89436142 \,, 
\end{equation}
to transform the data given in \cite{Filinov23} into atomic units, $r_s=\left[3/(4 \pi n a_{\rm Bohr}^3)\right]^{1/3}$ is dimensionless.

\begin{figure}[htp]
\centerline{ \includegraphics[width=0.5 \textwidth]{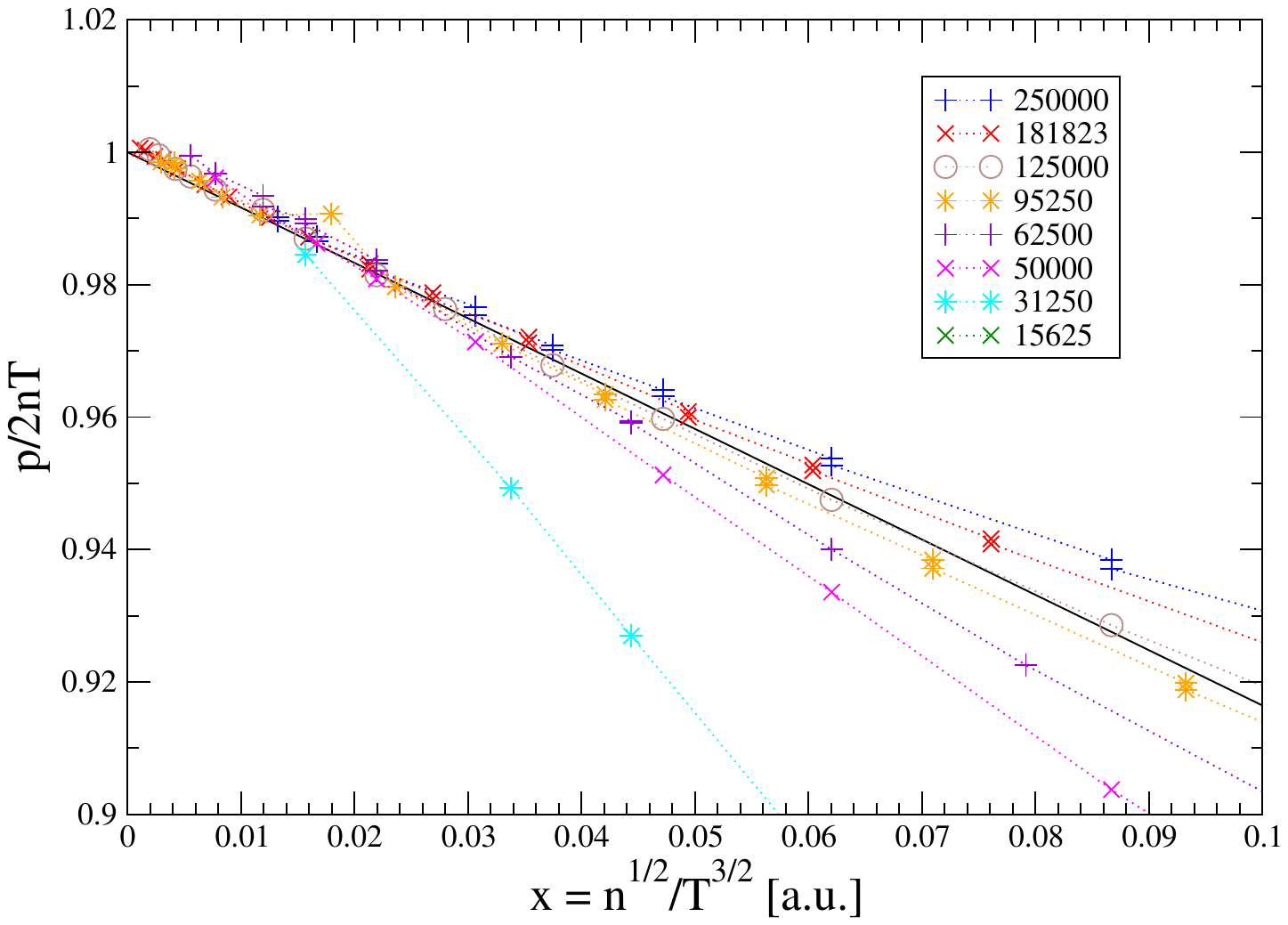} \includegraphics[width=0.5 \textwidth]{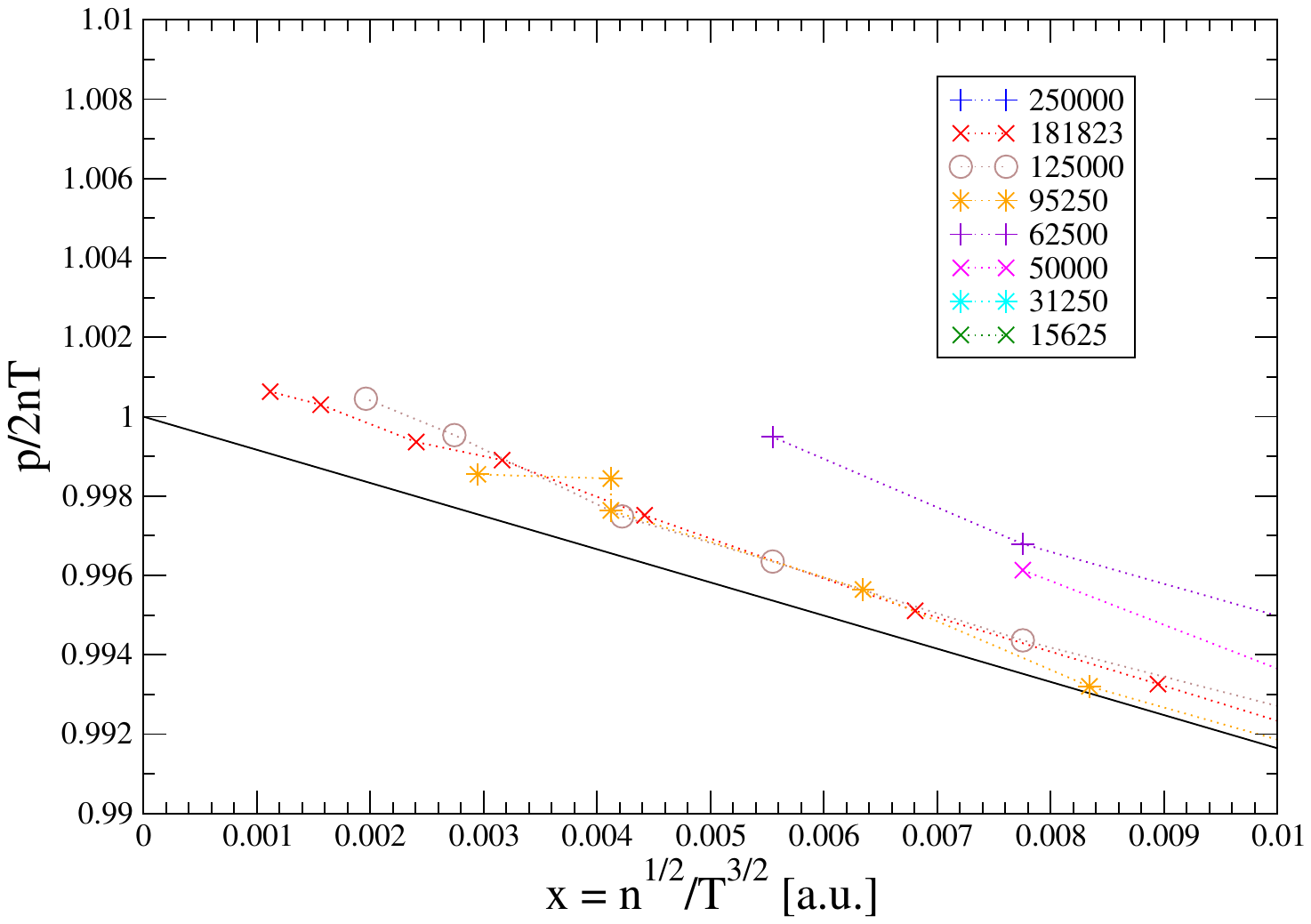}}
\caption{same as Fig. \ref{fig:1} with reduced $x$ ranges. \label{fig:1b}}
\end{figure}

The data shows only a slight scattering around a fairly smooth curve, so that the accuracy is high, the statistical errors given in the tables seem to be adequate in this respect. 
In the low-density limit, the isotherms approach the Debye limit within 1\% deviation for  $x < 0.04$ and  $T \ge 50\,000$ K ($T_{\rm Ha} \ge 0.15$), see Fig. \ref{fig:1b}. 
The low-temperature isotherm $T= 31\,250$ K shows strong deviations already at $x=0.02$, whereas the lowest isotherm $T=15\,625$ K does not approach the Debye limit within the tabulated set of data. 
The deviation from the Debye behavior indicate the relevance of the next virial term $ {\cal O}[n] $ to be discussed in the following section. 
In particular, this next virial coefficient describes also the formation of bound states, the Hydrogen atom. 
With decreasing $T$, the isotherms approach  the value $\beta p/(2n) =1/2$, which is valid for the atomic Hydrogen gas, at higher densities, see  Fig. \ref{fig:1}.

We now consider the Debye limit  in even more detail, see  Fig. \ref{fig:1b}.
Whereas the virial coefficient $A_{\rm Debye}$ is well reproduced for low densities, the simulation results deviate significantly from the lowest virial coefficient $A_{\rm ideal}$ if we look at the limit $x \to 0$ .
Deviations are  much larger than the statistical error of the simulations.

The pressure $p^{\rm PIMC}$ calculated in Ref. \cite{Filinov23} at the lowest densities seems to be somewhat too high.
A reason may be the use of the constants, $\tilde a_{\rm B}\approx 0.529$ \AA\, and   $\tilde k_{\rm B} \approx 1.38 \times 10^{-16}$ erg/K in the PIMC calculations \cite{Filinov23} with an accuracy much smaller than the required and stated $10^{-4}$.
The use of accurate values of the physical constants is necessary to obtain high-precision results for the equations of state.
For the lowest density, the deviations are even increasing. Possibly the finite size and the corresponding large screening length is problematic. 

\subsection{The effective second virial coefficient}

We now investigate  the second virial coefficient $A_2(T)$ according to Equ. (\ref{A2eff})   using the  exact expressions, see  (\ref{virialp}).  
 We consider isotherms $T=$ const. and calculate the effective, density-dependent second virial coefficient  using the results $A_{\rm ideal}(T)=1$ and  $A_1=0$, 
\begin{eqnarray}
\label{v2eff}
A_2^{\rm eff}(T,n)&=&-\frac{1}{n_{\rm Bohr}}\left[\frac{\beta p}{2 n}-1+A_{\rm Debye}(T)n_{\rm B}^{1/2}\right].
\end{eqnarray}

\begin{figure}[t]
\centerline{\includegraphics[width=0.6 \textwidth]{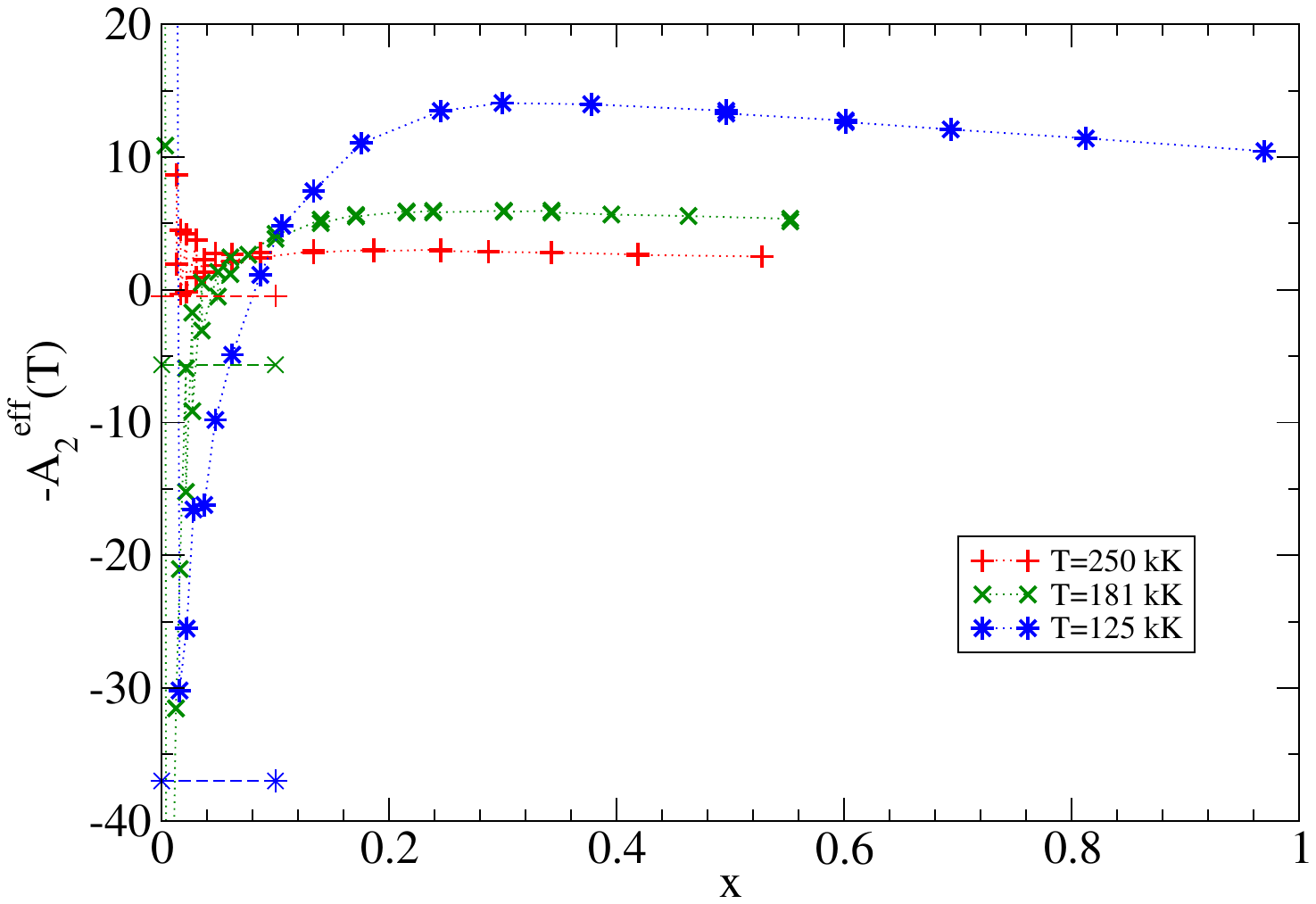}}
\caption{The effective second virial coefficient $A^{\rm eff}_2$ as function of $x=n_{\rm B}^{1/2}/T_{\rm Ha}^{3/2}$ for different temperatures.
Values for $A_2(T)$ according Tab. \ref{Tab:A2}  are shown as short dashed lines. 
\label{fig:2vir}}
\end{figure}

To show the zero-density limit for $A_2(T)= \lim_{n \to 0} A_2^{\rm eff}(T,n)$, we give a plot with the variable $x=n_{\rm B}^{1/2}/T_{\rm Ha}^{3/2}$ as used already in the Figs. \ref{fig:1} and \ref{fig:1b}. Results are shown in Fig. \ref{fig:2vir} together with the theoretical values from the exact expressions, see Tab. \ref{Tab:A2}.

There is  large scattering of $A_2^{\rm eff}(T,n)$ for $x \le 0.1$ so that a limit for $x \to 0$ can not be deducted. 
Possible reasons are the inaccuracy of the data, for instance due to finite size effects, what is seen in the large scatter of data. 
The expected values of $A_2(T)$ are given in Tab. \ref{Tab:A2}. 
Even for the highest temperature, these  data for $A_2(T)$ cannot be extracted from the PIMC simulations.

The virial plot to determine the subsequent virial coefficients $A_3(T), A_4(T)$ is shown in Appendix \ref{app:vir3}.
However, the accuracy of the current PIMC simulations is not sufficient to yield accurate results for these coefficients.

\subsection{Bound states and Saha equation}
\label{sec:BoundSaha}
At low temperatures, bound states become relevant. The bound state contribution $A^{\rm bound}_{2,ep}(T)$ to the second virial coefficient becomes dominating for $k_BT/E_{1s} \le 1$, see Tab.\,\ref{Tab:A2}.
If stronger bound states such as H$_2$ molecules are taken into account, the corresponding virial coefficient ($n_{\rm B}^4$) becomes the leading contribution to the pressure.
In this work, we discuss  only the formation of the hydrogen atom, not considering higher complexes.
Since the PIMC simulations \cite{Filinov23} are performed in the interesting range $k_BT/E_{1s} \approx 1$,
we study the role of bound state formation within the virial expansion.
Of particular interest are the low-temperature isotherms. In that case, 
 there is a problem because of the nonanalytical behavior $e^{E_{\rm bind}/T}$ at $T=0$ of the bound state contribution to the second virial coefficient, see  Eq.\,(\ref{PBL}) and text below.

\begin{figure}[htp]
\centerline{\includegraphics[width=0.4 \textwidth]{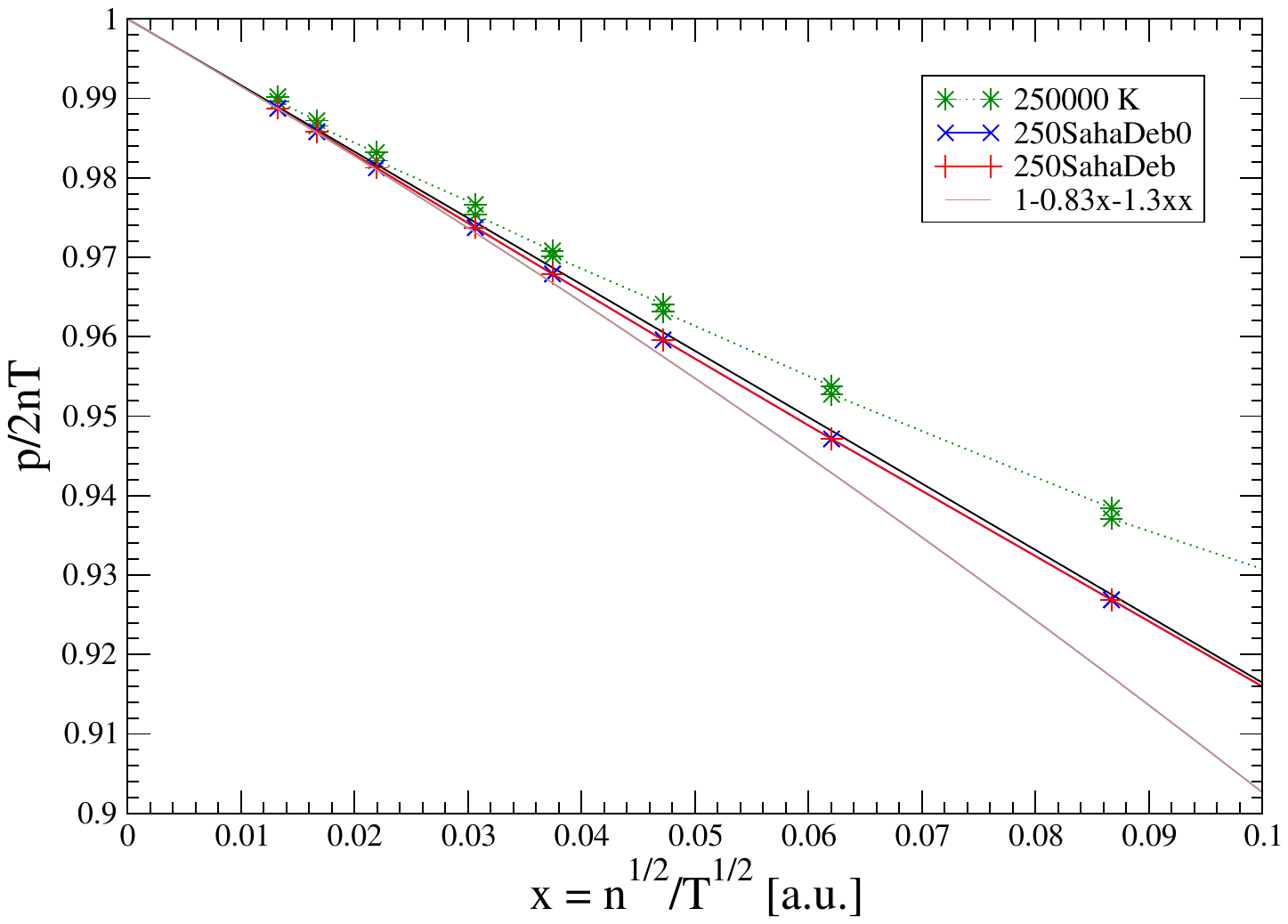}
\includegraphics[width=0.4 \textwidth]{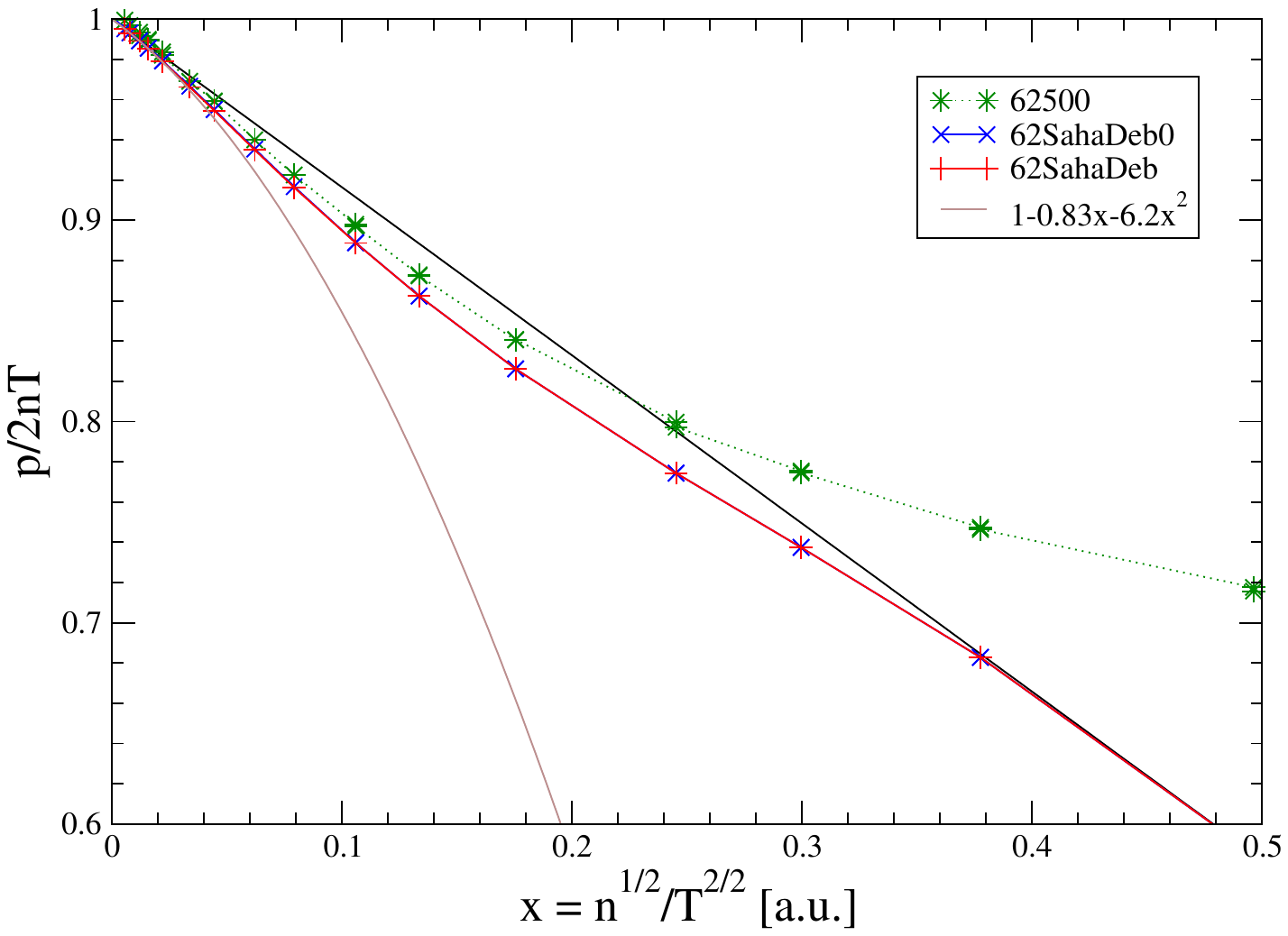}}
\label{fig:5}
\end{figure}

\begin{figure}[htp]
\centerline{\includegraphics[width=0.4 \textwidth]{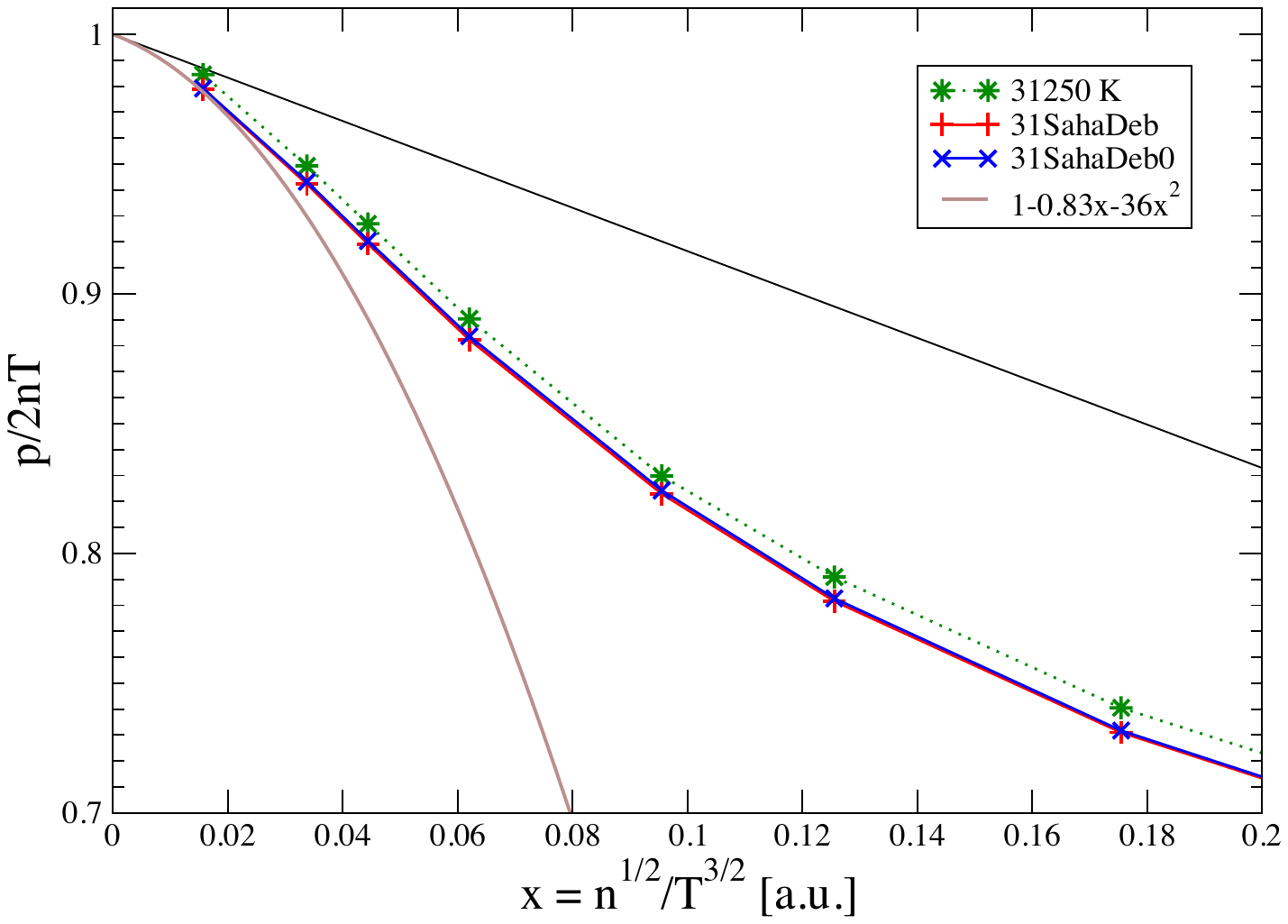} \includegraphics[width=0.44 \textwidth]{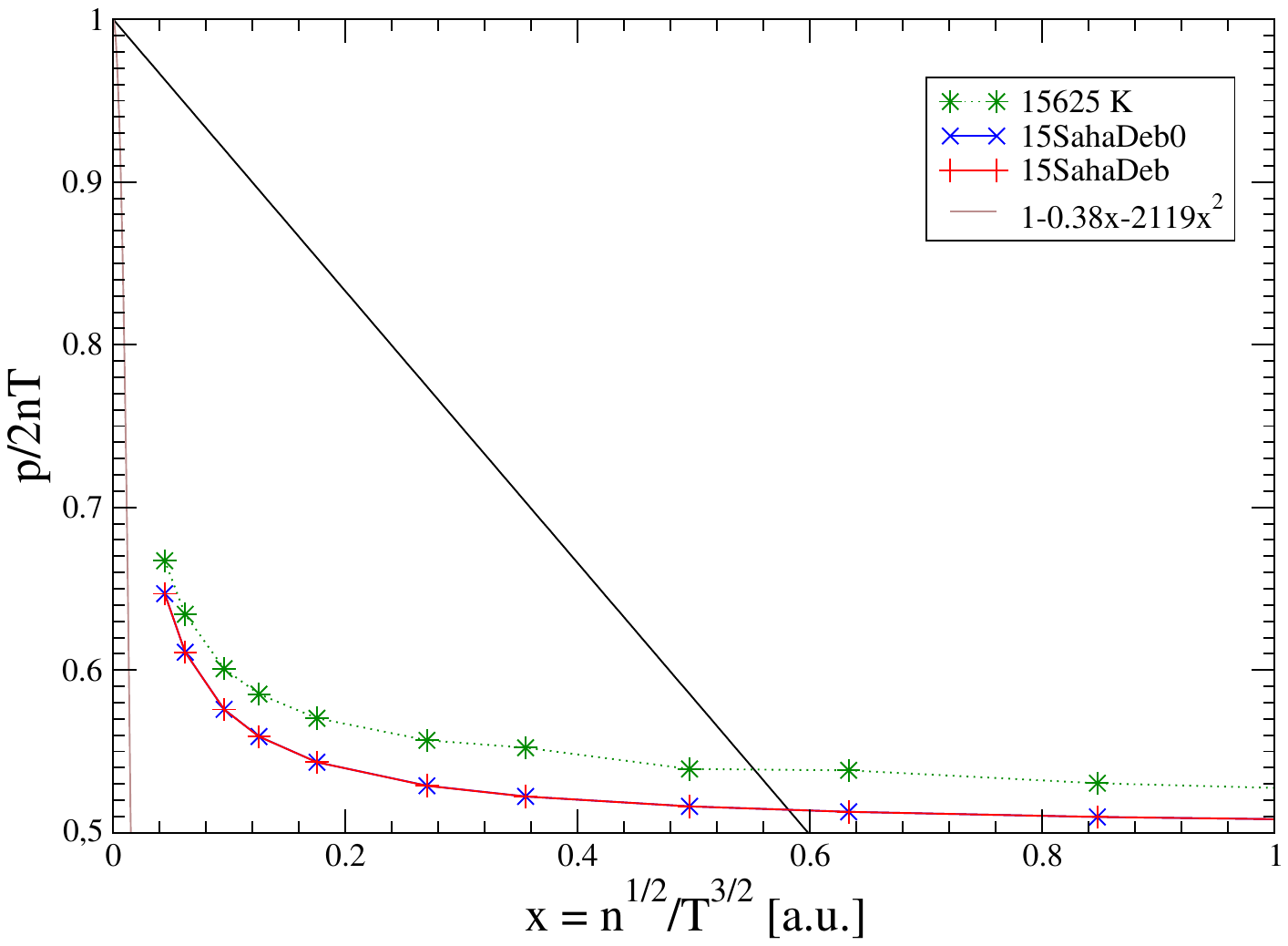}}
\caption{$\beta p/(2 n)$ as function of $n_{\rm Bohr}^{1/2}/T_{\rm Ha}^{3/2}$ at different temperatures: PIMC data \cite{Filinov23} (green *), Debye limit Eq.\,(\ref{pdebye}) (black full line),  virial Eq\,(\ref{virialSD0}) (pink full line), Planck-Brillouin-Larkin partition function: only ground state (SahaDeb0), all possible excited states (SahaDeb). 
\label{fig:31D}}
\label{fig:4}
\end{figure}


We show in Figs. \ref{fig:4} the virial plot for different temperatures. Beside the PIMC simulations we show the virial expansion up to first order, the Debye limit Eq.\,(\ref{pdebye}), as full black line.   The virial expansion up to the second order, Eq.\,(\ref{virialexp}) with Eqs.\,(\ref{virialp}) and (\ref{Qsigma})  has been taken in the following approximation
\begin{equation}
\label{virialSD0}
 \frac{\beta p}{2n}=1-\frac{(2 \pi)^{1/2}}{3}x- \frac{(2 \pi T_{\rm Ha})^{3/2}}{2} \left[e^{1/(2 T_{\rm Ha})}-1-\frac{1}{2 T_{\rm Ha}} \right]x^2,\qquad x=\frac{n_{\rm Bohr}^{1/2}}{T_{\rm Ha}^{3/2}}  
\end{equation}
where the last term is the bound part and has been taken explicitly from expanding the solution of the Saha equation (\ref{pTm}) with (\ref{saha0a}) for the ground state of the hydrogen atom. This is shown as light purple full line and the temperature dependent prefactor in front of $x^2$ is given explicitly in the legend box of each figure (-1.38114 for $T=250\,000$ K, -6.22667 for $T=62\,500$ K, -36.8587 for $T=31\,250$ K, -2119.19 for $T=15\,625$ K). Furthermore, the results from the full Saha equation  with the PBL partition function for the ground state  (\ref{pTm}) [SahaDeb0] and the PBL partition function for all possible states (\ref{PBL1}) [SahaDeb] are shown.

As expected, the PIMC simulations are well described by the virial expansion for low densities and high temperatures. The lower the temperature the more relevant is the improvement of the Debye approximation via the second virial coefficient. The deviation from Debye straight line starts at rather low densities and increases with lower temperatures. However, at the same time, the range where we find agreement of the simulations with the virial expansion up to the second order is limited to lower densities. At low temperatures, the virial expansion fails completely, see in particular the graphs for the lowest temperature of 15\,625\,K. The reason is the appearance of bound states. 

Obviously, the description using the Saha model for a partially ionized plasma works much better in the parameter range where bound states are relevant. Whereas SahaDeb0 takes only the ground state into account, SahaDeb considers  all bound states.
However, the contribution of the excited states is very small, in particular due to the renormalization terms $-1 +\beta E_s$ in the PBL partition function (\ref{PBL1}) or the corresponding term
in Eq. (\ref{alphaDeb}), so that no differences between SahaDeb and SahaDeb0 are seen in the figures. 

We see also that the PIMC data give a higher pressure than the Saha-Debye approximation.
The difference is larger than the statistical error of the calculation.
If the PIMC data have the accuracy as claimed by the authors, a possible explanation would be that the binding energy is reduced.
Then, the concentration of the free particles becomes higher so that the pressure is also higher.

\begin{figure}[t]
\centerline{
\includegraphics[width=0.6 \textwidth]{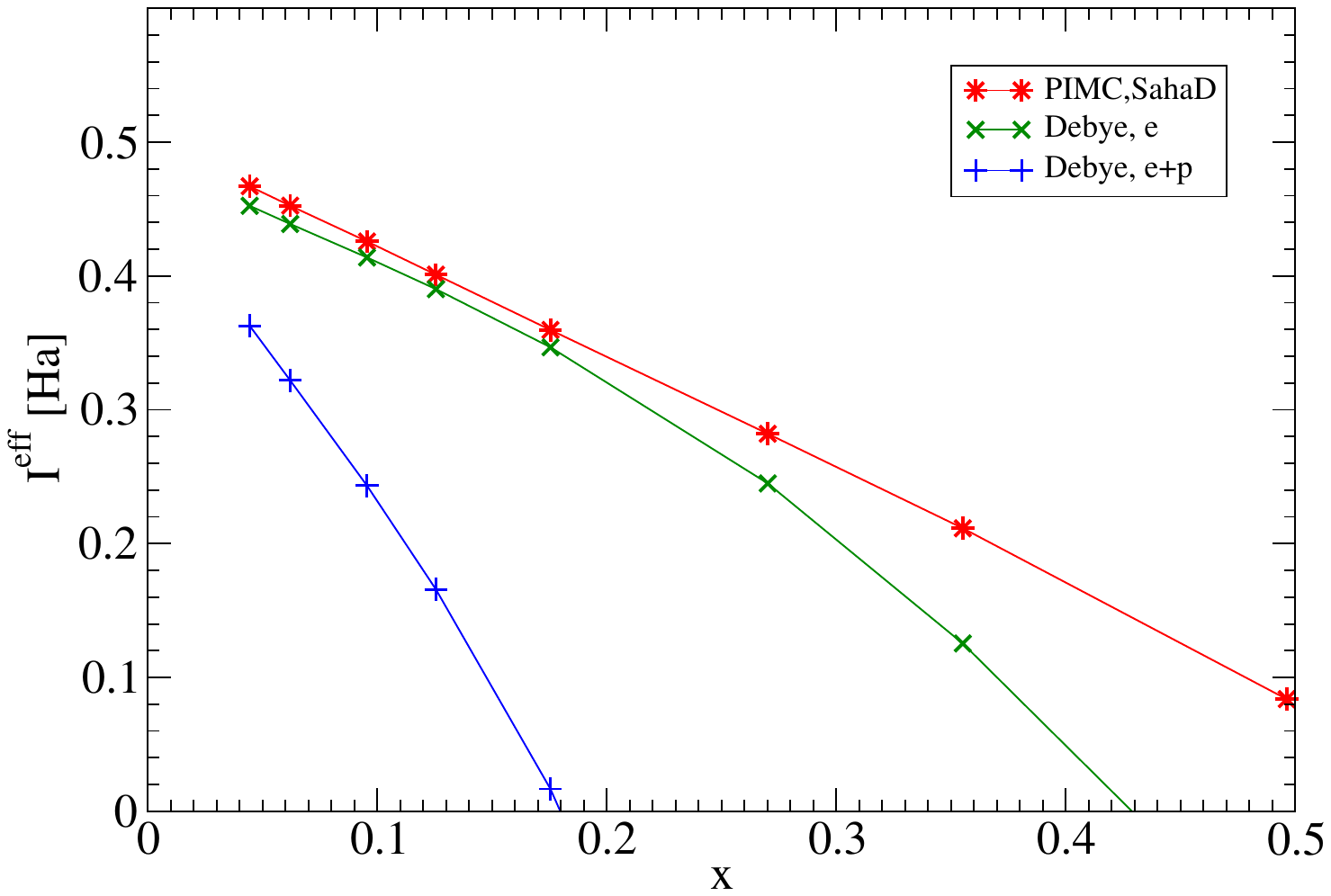}
}
\caption{The effective ionization energy $I^{\rm eff}_{\rm Ha}(T,n)$ as function of $x=n_{\rm B}^{1/2}/T_{\rm Ha}^{3/2}$. Isotherms for $T= 15\,625$ K, $T_{\rm Ha}=0.0494814$.
The Saha-Debye expression  $I^{\rm Saha0}_{\rm Ha}(T,n)$ (\ref{eq:Ieffec0}) (PIMC,SahaD) is obtained from the PIMC data.
The effective ionisation potential $0.5-2 \Delta^{\rm Deb}_{\rm Ha}$, Eq. (\ref{ISahaDeb}) (Debye,$e+p$, '+'), and $0.5- (\pi \alpha_e n_{\rm Bohr}/T_{\rm Ha})^{1/2}$ (Debye,$e$, 'x') are also shown.
\label{fig:IPD}}
\end{figure}

We try to extract the effective ionisation potential from the PIMC data, using the Saha-Debye approximation.
To analyse this, we introduce $I^{\rm eff}(T,n)$ defined according Eq. (\ref{eq:Ieffec}), see Fig. \ref{fig:IPD}.
The effective ionization energy is reduced what is known as IPD.
For comparison, we rewrite Eq. (\ref{SahaDeb}) replacing $(2/V) \sum_k e^{-\beta(E_k^{\rm qu}-\mu_e)}=n_e^*=n_p^*$.
For the quasiparticle shift in $E_k^{\rm qu}$ we take the Debye term (\ref{debyeshift}). After eliminating the chemical potentials, we introduce the ionization potential (see also (\ref{ISahaDeb}))
\begin{equation}
\label{ionizPot}
    I^{\rm Deb}(T,n)=-E^{\rm qu}_{10}-2 \Delta^{\rm Deb}(T,n).
\end{equation}
The shift of the ground state energy in $E^{\rm qu}_{10}$ is smaller than the value at $T=0$ which is $ 6 \pi T_{\rm Ha}^3 x^2=0.00228 x^2$ Ha for $T=15625 \,{\rm K} = 0.0494814 \, {\rm Ha}$ according to Eqs. (\ref{eq:Fockb}), (\ref{eq:Pauli}).
The Fock and Pauli blocking contributions to the bound-state quasiparticle energy $E_{10}^{\rm qu}$ gives no visible effect in the non-degenerate case.
It remains the Debye shift Eq. (\ref{ISahaDeb}), which is shown in Fig. \ref{fig:IPD}. 
The ionisation potential depression to describe the PIMC data is smaller than the Debye shift which takes both electrons and protons into account. 
For comparison, we calculated the shift only due to the electrons, i.e. screening only by the electrons and the lowering of the edge of the continuum only by one component,
$0.5- (\pi \alpha_e n_{\rm Bohr}/T_{\rm Ha})^{1/2}$, where consistently $\alpha_e$ is calculated with screening by electrons. 
The reasonable coincidence with the PIMC data can be considered as a hint to include dynamical screening and structure factor effects,
what is beyond the scope of the present work.


The data analysed in this work refers to the nondegenerate case,
effects of degeneracy occur for densities above a critical value $x_{\rm deg}^2=n^{\rm deg}_{\rm Bohr}/T_{\rm Ha}^3=2/(2 \pi T_{\rm Ha})^{3/2}$. 
At low temperatures, a further reduction of the pressure with increasing density is expected owing to the formation of H$_2$ molecules  \cite{Filinov23}. This issue is not discussed in our work.\\


\section{Discussion}
\label{sec:discussion}

\subsection{Virial expansion}
The validity of the virial expansion (up to the second virial coefficient) has been analysed in this work. It is limited to low densities and high temperatures. 
As $T$ decreases, the bound state  part becomes dominant. The virial expansion can only be applied at very low densities, whereas the Saha-Debye model already provides a reasonable description of the PIMC data at low temperatures ($T_{\rm Ha} < 1$) over a wide density range.

In this work, we propose an alternative form for describing the dependence of thermodynamic functions on density.
Based on the generalized Beth-Uhlenbeck formula (\ref{BUmedium}), the density expansion for the energy shifts of the quasiparticles should be performed.
The self-energy shifts of the single-particle states have been well studied. 
If the bound states are considered as quasiparticles, the energy shifts can be obtained.
The medium modifications of the continuum states are notoriously difficult to calculate. 
One possible solution to this problem is to use separable potentials, see Ref. \cite{Schmidt1989}, where medium-dependent scattering lengths and effective ranges can be derived.

The PIMC simulations are not yet perfect enough to determine higher orders of the virial expansion.
The accuracy should be improved and should be extended to higher densities.
Size effects due to the finite number of particles must be controlled and the sign problem must be addressed. 
Now that PIMC data for the UEG have been published \cite{TD,R24}, which show high accuracy, it would be important to provide such highly accurate results for hydrogen plasma as well.
The behavior near the Mott density would be of interest.

\subsection{Ionization potential depression (IPD)}

The ionization potential is the minimum energy required to excite a bound state so that an electron can escape (continuum of scattering states).
Experimentally, it determines the optical spectrum of the plasma and is related to the frequency-dependent dielectric function.
For dilute hydrogen plasma, it is a two-particle property and is given in the zero-density limit by solving the Schrödinger equation for the hydrogen atom. 
The ionization potential determines the various contributions in the Beth-Uhlenbeck formula (\ref{SahaDeb})
or the composition in the Saha equation, see (\ref{ionizPot}).

The definition of the ionization potential becomes problematic when the atom is embedded in a medium.
In Sect. \ref{sec:mediumTwo}, we presented an in-medium Schr\"odinger for the hydrogen atom, Eq. (\ref{imSGl}),
which contains the self-energy shifts of the single quasi-particle states and the shift of the bound states.
However, due to collisions in the plasma, the states also have a finite lifetime, and the energy levels are broadened.
The shift and broadening of the spectral lines has been well studied, see \cite{Guenter}.
The free quasiparticle states are also broadened.
Consequently, there is no sharp edge of the continuum of scattering states (i.e., the free states), but with increasing density, the spectral lines are washed out and disappear in the spectrum (Inglis-Teller effect).

To define the ionization potential in a dense medium, we consider only the real part of the self-energy, which determines the shifts of the continuum states. 
We also consider the shifts of the energies of the bound states, neglecting the broadening.
The change in the ionization potential caused by these shifts is referred to as the ionization potential depression (IPD).
The bound states disappear when the shifted energy of the bound state matches the corresponding quasi-particle energies of the electron and the ion, measured at the same center-of-mass momentum.
The bound state merges with the continuum (Mott effect).
At high densities and low temperatures, the Pauli blocking is important for the bound states, but also for the free states, since a transition to the continuum requires that the final states be free, for example above the Fermi energy for $T=0$.

In a certain approximation, the results for the shift of continuum states were considered in Sect. \ref{sec:mediumSingle},
and for bound states in Sect. \ref{sec:mediumTwo}.
The results for the IPD were discussed in \cite{Lin17,Lin2, Lin19}. 
In particular, it was shown that the ion-ion structure factor is important for the shift of the continuum states.
In the degenerate case, the Pauli blocking is crucial for the shift of the energies of the bound states and, at a critical density, for the merging with the continuum states.

A more detailed discussion of the IPD in connection with PIMC simulations was given recently by Bonitz and Kordts \cite{Bonitz25}.
The subdivision of electrons into bound and free electrons employing the trajectories of the electrons obtained from PIMC simulations \cite{Filinov23} is not free from arbitrariness.
In Ref. \cite{Bonitz25}, the mass-action law was discussed and the chemical potentials were determined.
The spectral function and the level shifts were presented, what is also studied in our work.
Two different approaches are used to analyze the PIMC data, with and without the bound state level shifts.
Results for the effective ionization energy of the ground state are shown.
The Mott density was inferred by extrapolating the IPD to higher densities.
The use of PIMC simulations can help to give a better understanding of concepts such as IPD in a dense plasma.
The medium modifications of energy levels, should be related to spectroscopic properties.
PIMC simulations of the polarisation function can give detailed information about the quasiparticle properties in dense plasmas.


\subsection{Ionization degree} 
\label{sec:ionization}

A characteristics of a plasma is the existence of free charge carriers with density $n^*$ which are relevant for conductivity.
We require a physical property to define the density of free electrons.
The electrical conductivity can be expressed in terms of a free carrier density and a collision frequency, but this gives no definition of the free carrier density from the observed conductivity.
We have in disordered structures the subdivision of electron states in localised and delocalized states, 
defining the edge of mobility.
Hopping processes and the Mott minimum conductivity have been considered in this context.

An important phenomenon for Coulomb systems is screening, which appears also as leading term in the virial expansion.
Not considering QED effects, it is a strict result for plasmas, described by the Coulomb potential.
The transition to a Debye potential is given by the free charged particles. The bound electrons contribute to the polarizability of the matter, but will not regularize the Coulomb divergence.

For a systematic approach, we consider the dielectric function 
\begin{equation}
    \epsilon(k,\omega)=1- \frac{e^2}{\epsilon_0k^2} \Pi(k,\omega).
\end{equation}
The polarization function (which is related to the dynamical structure factor or the density response function)
can be expressed in terms of Feynman diagrams \cite{RD79}, and two-particle properties are systematically included.
We consider the expansion in the long-wavelength limit
\begin{equation}
\label{polfct}
\Pi(k, \omega=0) = \epsilon_r \epsilon_0 e^{-2} \kappa_{\rm free}^2+(\epsilon_r-1)\epsilon_0e^{-2} k^2+ {\cal O}(k^4).      
\end{equation}
The first term with $\kappa_{\rm free}$ is determined by the free quasiparticle density, whereas the second term contains via $\epsilon_r$ the bound states (local field correction, Clausius-Mossotti term).
In the low-density, nondegenerate limit considered here, we have with $\epsilon_r \approx 1$
\begin{equation}
    \kappa^2_{\rm free}=\frac{2 e^2 n^*}{\epsilon_0 k_BT}.
\end{equation}

Within a general approach to the dielectric function, see \cite{Reinholz05,Guenter},
the polarization function which is related to the dynamical structure factor is related to the spectral function $A(k, \omega)$.
Within our approach, the free electron density is defined by the single quasiparticle contribution to the total electron density,
\begin{equation}
    n_e^*(T, \mu_e)=\frac{1}{V} \sum_kf_e(E_k^{\rm qu})
\end{equation}
with the quasiparticle energy defined by the single-particle peak in the spectral function. 
We considered different contributions to the self-energy which determine the shift of the quasiparticle energy, see Sect. \ref{sec:mediumSingle}.  
For the relation to the dynamical structure factor see Ref. \cite{Landau18}.
However, this definition anticipates a clear separation of the quasiparticle peak in the spectral function what is not always possible.

A similar approach was discussed recently \cite{Bellenbaum2025}. Instead of considering the dielectric function, the electronic response function was considered and the Chihara decomposition was taken to compare with a chemical picture.
PIMC simulations are used in a limited range of $q$ space to perform an average. 
They propose to fit the chemical dynamical structure factor to PIMC data.
The $e-e$ dynamical structure factor is decomposed into the elastic (Rayleigh) part, which describes the contribution of electrons following the ion motion, and an inelastic part which describe scattering processes of bound and free electrons.
This allows us to introduce the ionization degree, which is calculated from the best fit between the PIMC results and the chemical model.
This approach avoids the use of models like the Saha equation to derive either IPD or ionization state.

An interesting approach to calculate the ionization state of a material was presented in \cite{Mandy20}.
The optical conductivity $\sigma(\omega)$ was considered. 
Applying the Thomas-Reiche-Kuhn sum rule, the integral over the optical conductivity is related to the total number of electrons,
\begin{equation}
    n_e=\frac{2 m_e }{\pi e^2}\int_0^\infty d \omega \,\sigma(\omega).
\end{equation}
Within the Kubo-Greenwood approach, using single-electron states according to the DFT formalism, the dynamical (optical) conductivity can be decomposed into bound-bound, bound-free, and free-free contributions. 
Then, the free electron density can be defined as the contribution of free-free transitions to the sum rule.
The problem is that we have not a sharp gap without electron states since the system is disordered, so that the subdivision into different contributions is not strict. 
In addition, correlations are neglected in the DFT approach what makes the treatment of different ionization states for the ions in the plasma not exact.

A semiempirical approach to the ionization degree is given by the average atom model.
Atomic levels are calculated within a ion sphere, and the broadening of the levels to a band structures can be modelled, see, e.g.,
Chabrier et al. \cite{Chabrier05}.
The average atom model has been worked out recently by Zeng et al. \cite{Zeng20,Zeng22,Huang24,Zeng25}. 
Excitations in complex atoms, valence-band-like structure in a generalized ion-sphere approach with states which are either bound, free, or mixed states are considered. 
The Pauli principle is semiempirically taken into account by the ion sphere with a radius given by the density. 
It has been applied to various materials, traditional models such as the Ecker-Kr\"oll model have been improved.
Improvements of the average atom model and the corresponding IPD are performed in \cite{Luo2025} where a two-temperature approach was considered. 
While these semi-empirical models describe properties such as IPD and ionization state very efficiently, comparison with analytical results and PIMC simulations can help to give them a better foundation.

\section{Conclusions}\label{sec:Concl}

As for the UEG, PIMC simulations for the H plasma can be performed with high accuracy \cite{Millitzer,Filinov23}. Comparison with analytical approaches shows the validity of the virial expansion, but the current accuracy of PIMC is not sufficient to extract higher-order virial coefficients. 

Based on a Beth-Uhlenbeck approach, a quantum statistical approach to the equation of state is given. 
The convergence range of the virial expansion is reduced at low temperatures when bound states occur. 
We propose a quasiparticle expansion with medium-modified single- and two-particle properties. 
Quasiparticle shifts of free electrons and protons as well as of bound states are taken into account, and a virial expansion is possible. 
The treatment of continuum states remains a challenge.

The definitions of ionisation potential depression and ionisation state are currently under discussion.
It is important to relate these concepts to physical quantities that are free of arbitrariness. The behavior of the spectral function (self-energy shift of free and bound states) and the dielectric function (long-wave limit) are considered as possible concepts to derive definitions for these quantities.

\section*{Author contributions}

This is an author contribution text. 

\section*{Acknowledgments}
The authors thank Michael Bonitz, Tobias Dornheim, Jan Vorberger, and Wolf-Dietrich Kraeft for discussions.

\section*{Financial disclosure}

The authors recieved no specific for this work.

\section*{Conflict of interest}

The authors declare no potential conflict of interests.


\section*{Supporting information}

Additional supporting information may be found in the
online version of the article at the publisher’s website.

\appendix

\section{Higher virial coefficients}\label{app:vir3} 
\label{app:vir3}

We study the virial plot which provides us with the higher virial coefficients.
The density dependence of $A_2^{\rm eff}(T,n)$ in the low-density limit is given according Eq. (\ref{virialexp}) as
\begin{eqnarray}
\label{v2eff1}
A_2^{\rm eff}(T,n)&=&A_2(T)+A_3(T)n_{\rm B}^{1/2} \ln(n_{\rm B})+
A_4(T) n_{\rm B}^{1/2}+{\cal O}(n_{\rm B} \ln(n_{\rm B})\nonumber \\
&=&A_2(T)+A_3(T)n_{\rm B}^{1/2} \ln\left[B_4(T)n_{\rm B}\right]+{\cal O}(n_{\rm B} \ln(n_{\rm B})
\end{eqnarray}
with 
\begin{equation}
    A_4(T)=A_3(T) \ln [B_4(T)].
\end{equation}
Thus, in the virial plot where $A_2^{\rm eff}(T,n)$ is shown as a function of $y=n_{\rm B}^{1/2} \ln\left[B_4(T)n_{\rm B}\right]$, isotherms should meet the co-ordinate at $A_2(T)$, and the slope is $A_3(T)$. 
The coefficient $B_4(T)$ is obtained for the linear relation in this virial plot.
The linear pattern is violated when higher virial coefficients become relevant. Note that both, $A_2(T)$ and $A_3(T)$, are known for the Hydrogen plasma according Eq. (\ref{A2(T)}).

\begin{figure}[t]
\centerline{\includegraphics[width=0.6 \textwidth]{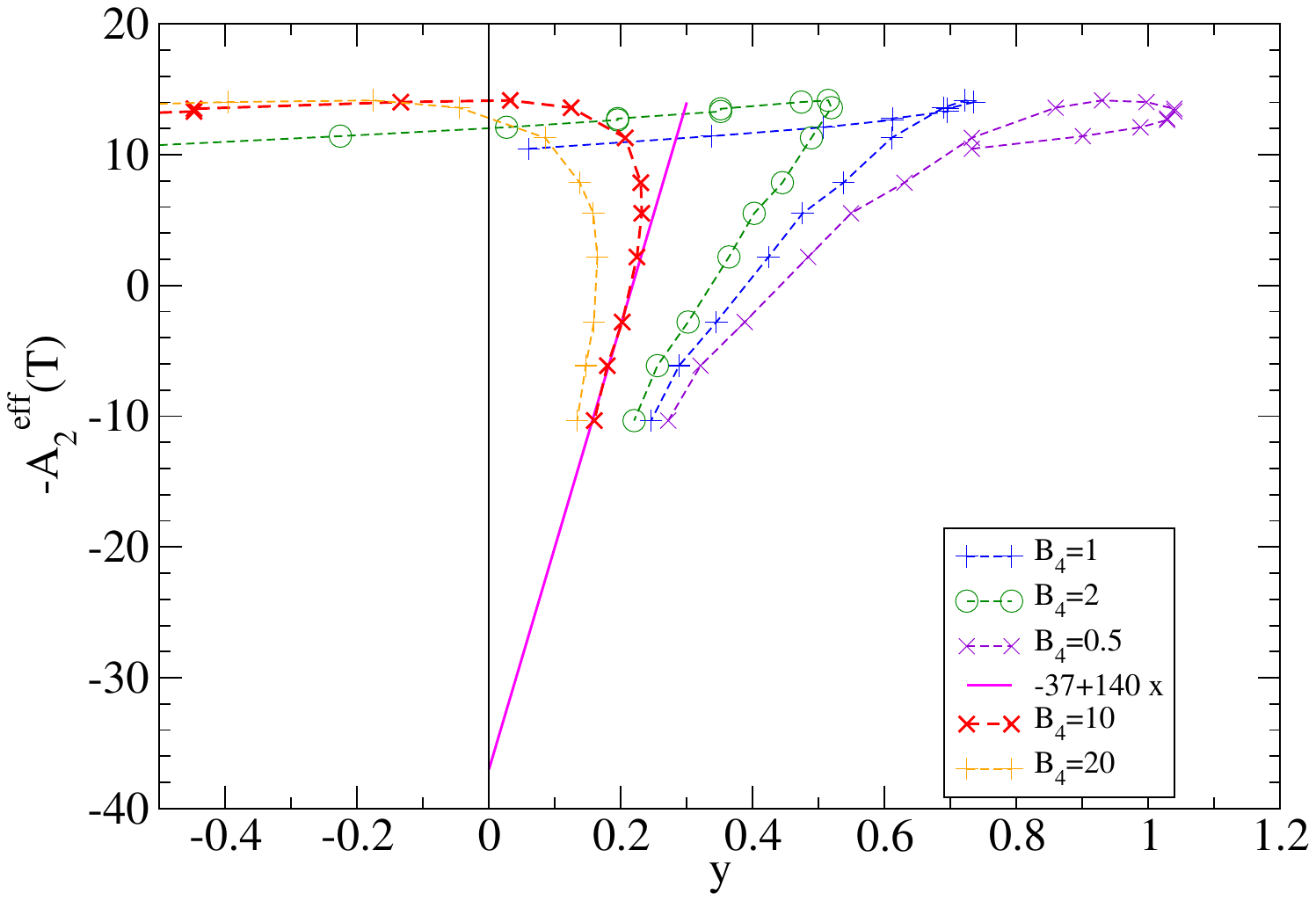}}
\caption{The effective second virial coefficient $A^{\rm eff}_2$ as function of $y=n_{\rm B}^{1/2} \ln\left[B_4(T)n_{\rm B}\right]$,
$T = 125000$ K.
\label{fig:3vir125}}
\end{figure}

We apply this method of the virial plot to PIMC simulations $p^{\rm PIMC}(T,n)$ to infer the virial coefficients $A_3(T), A_4(T)$ from  the values $A_2^{\rm eff,PIMC}(T,n)$ according Eq. (\ref{v2eff}).
As example we present results for the temperature $T=125000$ K, Fig. \ref{fig:3vir125}.
The values for the lowest densities show a strong scatter and are dropped.
It is obvious that for this method to extract virial coefficients, high precision of calculated data $p^{\rm PIMC}(T,n)$ is required. 
We are analysing the difference of big numbers, since the lower virial coefficients such as the Debye term dominate the low-density limit of the pressure. 
We consider $B_4(T)$ in the variable $y$ as a parameter. 
The best choice of $B_4$ is the graph where the linear relation is best realized.
Considering the different graphs in Fig. \ref{fig:3vir125}, the PIMC data follow a linear relation as long as possible
for $B_4 \approx 10$.

If we take the value of $-A_2(T=125000 \,{\rm K})=-37.004$ given in Tab. \ref{Tab:A2}, the best linear fit is
\begin{equation}
    -A^{\rm eff}_2(T_{\rm Ha}=0.395851)=-37.004+140 \,y= -37.004+140\, n_{\rm B}^{1/2} \ln\left[10 \, n_{\rm B}\right].
\end{equation}
According to the virial expansion (\ref{virialp}), the values of $A_3$ for $T_{\rm Ha}=0.395851$ is  509.744. 
The fitted parameter value $A_3^{\rm PIMC}\approx 140$ is too small. 
The fitted value $B_4 \approx 10$ results in a value $A_4 \approx  322$ for $T_{\rm Ha}=0.395851$.
 We only show the procedure here, but the fitted values are very rough due of the strong scatter of the PIMC data at low densities.

\end{document}